\pdfoutput=0
\documentclass[11pt]{article}
\usepackage{epsfig}
\usepackage{amsfonts}
\usepackage{amsmath}
\usepackage{bm,bbm}
\usepackage{cite}
\usepackage[hidelinks]{hyperref}
\allowdisplaybreaks[1]

\newcommand{\la}[1]{\label{#1}}
\newcommand{\be}{\begin{equation}}
\newcommand{\ee}{\end{equation}}
\newcommand{\ba}{\begin{eqnarray}}
\newcommand{\ea}{\end{eqnarray}}
\newcommand{\rmi}[1]{{\mbox{\scriptsize #1}}}
\newcommand{\fig}{Fig.~}
\newcommand{\figs}{Figs.~}
\newcommand{\eq}{Eq.~}
\newcommand{\eqs}{Eqs.~}
\newcommand{\se}{Sec.~}

\newcommand{\nr}[1]{(\ref{#1})}
\newcommand{\tr}{{\rm Tr\,}}
\newcommand{\nn}{\nonumber \\}
\newcommand{\fr}[2]{{\frac{#1}{#2}\,}}

\renewcommand{\vec}[1]{{\bf #1}}

\renewcommand{\eq}{eq.~}
\renewcommand{\eqs}{eqs.~}
\renewcommand{\se}{sec.~}

\renewcommand{\fig}{fig.~}
\renewcommand{\figs}{figs.~}

\newcommand{\rmO}{{\mathcal{O}}}
\newcommand{\bmu}{\bar\mu}

\def\lsi{\raise0.3ex\hbox{$<$\kern-0.75em\raise-1.1ex\hbox{$\sim$}}}
\def\gsi{\raise0.3ex\hbox{$>$\kern-0.75em\raise-1.1ex\hbox{$\sim$}}}

\newcommand{\sign}{\mathop{\mbox{sign}}}
\newcommand{\nF}{f_\rmii{F}} 
\newcommand{\nB}{f_\rmii{B}} 
\newcommand{\rmii}[1]{{\mbox{\tiny\rm{#1}}}}
\newcommand{\rmiii}[1]{{\mbox{\tiny{$\scriptstyle{\rm#1}$}}}}
\newcommand{\re}{\mathop{\mbox{Re}}}
\newcommand{\im}{\mathop{\mbox{Im}}}
\newcommand{\Tint}[1]{{\hbox{$\sum$}\!\!\!\!\!\!\!\int\,}_{\!\!\!\!\raise-0.9ex\hbox{$\scriptstyle{#1}$}}}
\newcommand{\Tinti}[1]{{{\Sigma}\!\!\!\!\raise0.3ex\hbox{$\int$}_\rmii{${#1}$}}}
\newcommand{\bi}{\begin{itemize}}
\newcommand{\ei}{\end{itemize}}
\newcommand{\hide}[1]{ }
\newcommand{\bsl}[1]{\,\slash\!\!\!\!{#1}\,}

\newcommand{\deltabar}{\raise-0.02em\hbox{$\bar{}$}\hspace*{-0.8mm}{\delta}}
\newcommand{\ddeltabar}{\raise-0.18em\hbox{$\bar{}$}\hspace*{-0.8mm}{\delta}}
\renewcommand{\P}{\mathcal{P}}

\newcommand{\K}{\mathcal{K}}
\newcommand{\Q}{\mathcal{Q}}
\newcommand{\X}{\mathcal{X}}

\newcommand{\A}{\mathcal{A}}
\newcommand{\B}{\mathcal{B}}
\newcommand{\C}{\mathcal{C}}
\newcommand{\D}{\mathcal{D}}

\newcommand{\F}{\mathcal{F}}

\newcommand{\iI}{\rmii{I}}

\newcommand{\T}{\rmii{$T$}}

\newcommand{\aW}{\rmii{$W$}}
\newcommand{\xW}{s^2_\aW}

\newcommand{\lnf}{l^{ }_\rmi{1f}}
\newcommand{\lif}{l^{ }_\rmi{2f}}
\newcommand{\ltf}{l^{ }_\rmi{3f}}
\newcommand{\fin}{\mbox{\sl f\,}}
\newcommand{\ini}{\mbox{\sl i\,}}
\newcommand{\sfin}{\rmii{\sl f\,}}
\newcommand{\sini}{\rmii{\sl i\,}}

\makeatletter \@addtoreset{equation}{section} \makeatother
\renewcommand{\theequation}{\arabic{section}.\arabic{equation}}
\makeatletter
\renewcommand\section{\@startsection {section}{1}{\z@}%
                                   {-5.5ex \@plus -1ex \@minus -.2ex}
                                   {2.3ex \@plus.2ex}%
                                   {\normalfont\large\bfseries}}
\renewcommand\subsection{\@startsection{subsection}{2}{\z@}%
                                     {-3.25ex\@plus -1ex \@minus -.2ex}%
                                     {1.5ex \@plus .2ex}%
                                     {\normalfont\normalsize\bfseries}}
\renewcommand\thesection {\@arabic\c@section}
\renewcommand\thesubsection   {\thesection.\@arabic\c@subsection}
\renewcommand{\@seccntformat}[1]{%
\csname the#1\endcsname.\hspace{1.0em}}
\makeatother

\begin{document}

\flushbottom

\begin{titlepage}

\begin{flushright}
May 2025
\end{flushright}
\begin{centering}
\vfill

{\Large{\bf
  $\nu\bar\nu$ production, annihilation, and scattering\\[2mm]
  ~at MeV temperatures and NLO accuracy
}} 

\vspace{0.8cm}

G.~Jackson$^\rmi{a}$ 
and
M.~Laine$^\rmi{b}$ 
% ...

\vspace{0.6cm}

${}^\rmi{a}_{ }${\em
SUBATECH,
Nantes Universit\'e, 
IMT Atlantique, 
IN2P3/CNRS, \\ 
4 rue Alfred Kastler, 
La Chantrerie BP 20722, 
44307 Nantes, France \\}

\vspace*{0.3cm}

${}^\rmi{b}_{ }${\em
AEC, 
Institute for Theoretical Physics, 
University of Bern, \\ 
Sidlerstrasse 5, CH-3012 Bern, Switzerland \\}

\vspace*{0.6cm}

{\em 
Emails: jackson@subatech.in2p3.fr, laine@itp.unibe.ch}

\vspace*{0.8cm}

\mbox{\bf Abstract}
 
\end{centering}

\vspace*{0.3cm}
 
\noindent
Interaction rates of neutrinos and antineutrinos within a QED plasma
determine the dynamics of their decoupling in the early universe. 
We show how to define the relevant double-differential production, 
annihilation, and 
scattering rates at NLO. Integrating over these rates with specific weights, 
other quantities from the literature can be
obtained, such as energy transfer rates, or a neutrino interaction 
rate. In the limit of massless electrons, 
we show that NLO corrections to the energy transfer rates 
are as small as those that enter the previously
determined neutrino interaction rate, and 
only have a small influence on the neutrino decoupling parameter, 
$N_\rmi{eff}\,$. For comparison, the influence of a finite electron 
mass is quantified at LO. Finally we provide a tabulation and fast
interpolation routine for all double-differential rates,
in order to allow for their use in non-approximate kinetic equations, which
may further reduce the systematic uncertainties of the Standard Model
prediction for~$N_\rmi{eff}\,$. 

\vfill


\end{titlepage}

\tableofcontents

%
\section{Introduction}

An important probe of early universe physics is provided by the effective
number of neutrino species, $N_\rmi{eff}$. Neutrinos decouple from the
QED plasma at temperatures 
$T^{ }_\gamma \sim 2$~MeV, when the rate of weak interactions
becomes smaller than the Hubble rate, 
$G_\rmii{F}^2 T^5_{\gamma} < H$, 
where $G^{ }_\rmii{F}$ is the Fermi coupling, 
$T^{ }_\gamma$ the photon temperature, and $H$ the Hubble rate. 
After this moment, the neutrinos stream freely. 
In contrast, 
the QED plasma experiences a non-trivial change 
at $T^{ }_\gamma < 0.5$~MeV, when 
electrons become non-relativistic. After this period --- notably, 
during big bang nucleosynthesis (BBN), 
at $T^{ }_\gamma\sim 0.1$~MeV, or photon
decoupling (CMB), at $T^{ }_\gamma \sim 0.3$~eV --- 
the energy densities in neutrinos
and photons stay fixed, and can be parametrized as 
\be
 \frac{ e^{ }_\nu }{e^{ }_\gamma}
 \; \equiv \; 
 \frac{7}{8}\biggl( \frac{4}{11} \biggr)^{4/3}_{ } N^{ }_\rmi{eff} 
 \;.
 \la{def_Neff}
\ee
The numerical factors have been chosen so that in an idealized 
non-interacting limit, $N_\rmi{eff} \simeq 3\,$. 
Because of interactions, the Standard Model prediction for
$N_\rmi{eff}$ is not~3 but slightly larger. 

On the other hand, $N_\rmi{eff}$ 
can also be extracted empirically, 
via the effect that  
the total energy density of radiation, $e_\rmi{rad}\,$,
has on BBN and CMB physics. 
The radiation energy density may receive a 
partial component from 
physics Beyond the Standard Model (BSM), 
\be
 \Delta e^{ }_\rmii{BSM} 
 \; \equiv \; 
 e_\rmi{rad}
 -
 e^{ }_\gamma
 -
 e^{ }_\nu
 \;, 
 \la{def_Delta_e_BSM} 
\ee 
thus contributing 
to the measured $N_\rmi{eff}\,$.
Therefore, $N_\rmi{eff}$ represents
both a test of cosmology based on the Standard Model, 
as well as a stringent constraint on exotic BSM scenarios. 

The Standard Model prediction for $N_\rmi{eff}$ has reached 
impressive precision in the meanwhile. 
It is based on the solution to complicated kinetic 
equations, incorporating many physical 
phenomena~\cite{Neffm2,Neffm1,Neff0,Neff4}. 
The kinetic equations are 
parametrized by a number of coefficients. 
On one hand, these describe
how the universe expands at MeV temperatures; the expansion changes 
the temperature and therefore drives the system out of equilibrium. 
On the other hand, the rate equations contain microscopic equilibration
rates, induced by elastic or inelastic scatterings, 
which tend to keep the neutrino ensemble
close to equilibrium, by redistributing 
the energy between the various degrees of freedom. 

Recently, QED corrections to the rate coefficients
have been estimated~\cite{cemp,rate,new}. The magnitude of these
corrections is important for establishing 
the accuracy of the Standard Model prediction for $N_\rmi{eff}$,  
and indeed there is a debate about the  
third decimal place~\cite{cemp,new}. 

A part of the difficulty in resolving the third decimal 
of $N_\rmi{eff}$ is that the rate
equations used in the various studies, and consequently the rate coefficients 
that have been estimated in the literature, are not identical.  
One of the estimates~\cite{cemp} relied on ``averaged'' kinetic
equations~\cite{mea2,mea}. Another estimate carried out a complete 
computation of one of the rate coefficients appearing in the full kinetic
equations~\cite{rate}, but the full equations contain more coefficients
than this single one. The third computation~\cite{new} only analyzed
one diagram among the full set of quite many NLO QED corrections. 

The purpose of the present study is twofold. On one hand, we show how
an ``integrand'' of the rate coefficient considered in ref.~\cite{rate}
is itself physical (meaning gauge independent and IR finite), and makes
an appearance in the full kinetic equations. We refer to this integrand
as a double-differential rate. 
This opens up the avenue
to a complete NLO QED computation of $N_\rmi{eff}$
via the solution of full kinetic equations. On the other hand, 
going in the opposite direction, we show how certain moments of the 
double-differential rate yield the same 
energy transfer rates that appeared in ref.~\cite{cemp}.
Therefore we can now carry out an unambiguous comparison of 
NLO QED corrections between refs.~\cite{cemp} and \cite{rate}. 

Our presentation is organized as follows. 
In~\se\ref{se:intuition}, we review what leading-order Boltzmann
equations can say about the rate coefficients that 
characterize the interactions of neutrinos and antineutrinos
with a QED plasma. Subsequently, in \se\ref{se:qm}, we show
that a key object appearing in the Boltzmann equations can also
be defined and computed in full thermal field theory. Given this
object, energy transfer rates are analyzed in \se\ref{se:energy}.
Numerical results for the full double-differential
rate are tabulated in \se\ref{se:tabulation}, 
and we conclude in \se\ref{se:conclu}.

%
\section{Overall picture according to Boltzmann equations}
\la{se:intuition}

Given that Boltzmann equations are the standard tool
for estimating the efficiency of various processes in cosmology, 
we start by discussing what they say about the interactions of
neutrinos and antineutrinos with a QED plasma. Even if the Boltzmann
equations themselves cannot be directly extended to the NLO level, 
this intuition helps in identifying physically relevant observables. 

%
\subsection{Pair production and annihilation}

Let us consider a Boltzmann equation 
for the phase space 
distribution $f^{ }_{\vec{k}^{ }_\nu}$ of a neutrino 
of spatial momentum $\vec{k}^{ }_{\nu}$.
Focussing first on 
leading-order pair creation and annihilation (``$s$-channel'')
processes, it can be written as  
\be
 \dot{f}^{ }_{\vec{k}^{ }_\nu}
 \supset
  \frac{1}{2k^{ }_\nu}
  \int^{ }_{\vec{p}^{ }_{e},\vec{p}^{ }_{\bar e},\vec{q}^{ }_{\,\bar\nu}}
  \hspace*{-11mm}
  {\rm d}\Phi^{ }_{
                   \vec{p}^{ }_{e}+\vec{p}^{ }_{\bar e} 
                   \to 
                   \vec{k}^{ }_{\nu}+\vec{q}^{ }_{\,\bar\nu}
                  }
 \Bigl[
 \underbrace{
  f^{ }_{\vec{p}^{ }_{e}}
  f^{ }_{\vec{p}^{ }_{\bar e}}
  (1 - f^{ }_{\vec{k}^{ }_\nu}) 
  (1 - f^{ }_{\vec{q}^{ }_{\,\bar\nu} }) 
 }_{\rm gain}
 - 
 \underbrace{
  f^{ }_{\vec{k}^{ }_\nu}
  f^{ }_{\vec{q}^{ }_{\,\bar\nu} }
  (1 - f^{ }_{\vec{p}^{ }_{e}}) 
  (1 - f^{ }_{\vec{p}^{ }_{\bar e}}) 
 }_{\rm loss}
 \Bigr]
 {\textstyle\sum}|\mathcal{M}|^2_{ }
 \;, \la{boltzmann_s}
\ee
where $k \equiv |\vec{k}|$, 
$
 \int_\vec{q} \equiv \int\! \frac{{\rm d}^3_{ }\vec{q}}{(2\pi)^3_{ }}
$, 
\be
  {\rm d}\Phi^{ }_{
                   \vec{p}^{ }_{e}+\vec{p}^{ }_{\bar e} 
                   \to 
                   \vec{k}^{ }_{\nu}+\vec{q}^{ }_{\,\bar\nu}
                  }
  \; \equiv \; 
 \frac{
  (2\pi)^4_{ }
  \delta^{(3)}_{ }( 
                    \vec{p}^{ }_{e} + \vec{p}^{ }_{\bar e}
                    - \vec{k}^{ }_{\nu} - \vec{q}^{ }_{\,\bar\nu} 
                   )
  \delta(
        \epsilon^{ }_{e} + \epsilon^{ }_{\bar e}
       -  k^{ }_\nu - q^{ }_{\,\bar\nu} 
         )
 }{
 (2 q^{ }_{\bar\nu})
 (2 \epsilon^{ }_{e})
 (2 \epsilon^{ }_{\bar e})
 }
 \;, 
\ee
and $\epsilon^{ }_e \equiv \sqrt{p^2_e + m^2_e}$.
On the left-hand side, 
$
  \dot{f}^{ }_{\vec{k}^{ }_\nu}
$
denotes a covariant time derivative in an expanding background. 

In order to define a {\em production rate}, we may envisage
an initial state in which the neutrino and antineutrino phase-space
densities have been set to zero. Of course, it should be stressed 
that this is just a thought experiment from the
cosmological point of view, however the coefficient 
defined, cf.\ \eq\nr{def_Psi}, is of broader applicability. 
In the said limit, 
the loss term drops out from \eq\nr{boltzmann_s}. 
As for the gain term, from which the 
Pauli blocking factors also drop out, 
it is helpful to factor out the integral over the momenta 
$\vec{p}^{ }_e$ and $\vec{p}^{ }_{\bar e}$, so we denote
\be
 \Psi(\vec{k}^{ }_{\nu},\vec{q}^{ }_{\,\bar\nu})
 \;\equiv\;
 \frac{1}{ 2 k_\nu }
  \int^{ }_{\vec{p}^{ }_{e},\vec{p}^{ }_{\bar e}}
  \hspace*{-5mm}
  {\rm d}\Phi^{ }_{
                   \vec{p}^{ }_{e}+\vec{p}^{ }_{\bar e} 
                   \to 
                   \vec{k}^{ }_{\nu}+\vec{q}^{ }_{\,\bar\nu}
                  }
  \
  f^{ }_\rmii{F} ( \epsilon^{ }_{e} )
  f^{ }_\rmii{F} ( \epsilon^{ }_{\bar e} )
  \;
 {\textstyle\sum}|\mathcal{M}|^2_{ }
 \;. \la{def_Psi}  
\ee
Here we have also put the electrons and positrons 
in thermal equilibrium, so that their phase-space densities
are given by Fermi distributions ($\equiv f^{ }_\rmii{F}$), 
characterized by a temperature~$T^{ }_\gamma$. 

With the help of \eq\nr{def_Psi}, 
\eq\nr{boltzmann_s} can now be expressed as
\be
 \dot{f}^{ }_{\vec{k}^{ }_\nu}
 \; \supset \;
 \int^{ }_{\vec{q}^{ }_{\,\bar\nu}}
 \Bigl[\,
 \Psi(\vec{k}^{ }_{\nu},\vec{q}^{ }_{\,\bar\nu}) 
  (1 - f^{ }_{\vec{k}^{ }_\nu}) 
  (1 - f^{ }_{\vec{q}^{ }_{\,\bar\nu} }) 
 - 
  f^{ }_{\vec{k}^{ }_\nu}
  f^{ }_{\vec{q}^{ }_{\,\bar\nu} }
 {\widetilde \Psi}(\vec{k}^{ }_{\nu},\vec{q}^{ }_{\,\bar\nu}) 
 \,\Bigr]
 \;. \la{b_with_f}
\ee
Here ${\widetilde \Psi}(\vec{k}^{ }_{\nu},\vec{q}^{ }_{\,\bar\nu})$
is the {\em annihilation rate}, which is related to the production rate 
by 
\be
 {\widetilde \Psi}(\vec{k}^{ }_{\nu},\vec{q}^{ }_{\,\bar\nu}) 
  \; \equiv \; 
  e^{(k^{ }_{\nu} + q^{ }_{\,\nu} )/T^{ }_\gamma}_{ }
  \; 
 {\Psi}(\vec{k}^{ }_{\nu},\vec{q}^{ }_{\,\bar\nu}) 
  \;. \la{balance_1}
\ee
Here we made use of 
\be
 [1 -  f^{ }_\rmii{F} ( \epsilon^{ }_{e} )]
 [1 - f^{ }_\rmii{F} ( \epsilon^{ }_{\bar e} ) ]
 \; = \;
 e^{( \epsilon^{ }_{e} + \epsilon^{ }_{\bar e}  )/T^{ }_\gamma}_{ }
  f^{ }_\rmii{F} ( \epsilon^{ }_{e} )
  f^{ }_\rmii{F} ( \epsilon^{ }_{\bar e} )
 \; = \;
 e^{(k^{ }_{\nu} + q^{ }_{\,\nu} )/T^{ }_\gamma}_{ }
  f^{ }_\rmii{F} ( \epsilon^{ }_{e} )
  f^{ }_\rmii{F} ( \epsilon^{ }_{\bar e} ) 
 \;.
\ee
The prefactor in \eq\nr{balance_1}
guarantees detailed balance: if 
$ f^{ }_{\vec{k}^{ }_\nu} \to f_\rmii{F}^{ }(k^{ }_\nu) $ and  
$ f^{ }_{\vec{q}^{ }_{\,\bar\nu} } \to f_\rmii{F}^{ }(q^{ }_{\bar\nu}) $, 
then the neutrino phase space densities satisfy
$
 [ 1 - f_\rmii{F}^{ }(k^{ }_\nu) ] 
 [ 1 - f_\rmii{F}^{ }(q^{ }_{\bar\nu}) ]
 =
  e^{(k^{ }_{\nu} + q^{ }_{\,\nu} )/T^{ }_\gamma}_{ }
 \,  
 f_\rmii{F}^{ }(k^{ }_\nu) 
 f_\rmii{F}^{ }(q^{ }_{\bar\nu})
$, 
implying that the right-hand side of \eq\nr{b_with_f} vanishes
in full equilibrium. 

\vspace*{3mm}

Now, taking moments of \eq\nr{def_Psi}, further quantities
can be defined. In particular, integrating over both momenta, yields
the idealized production rates of number densities, 
\be
 \dot{n}^\rmi{gain}_{\nu} \; \supset \; 
 \int_{\vec{k}^{ }_\nu,\vec{q}^{ }_{\,\bar\nu}} \hspace*{-1mm}
    \Psi(\vec{k}^{ }_{\nu},\vec{q}^{ }_{\,\bar\nu}) 
 \;, \quad
 \dot{n}^\rmi{gain}_{\bar\nu} \; \supset \; 
 \int_{\vec{k}^{ }_\nu,\vec{q}^{ }_{\,\bar\nu}} \hspace*{-1mm}
    \Psi(\vec{k}^{ }_{\nu},\vec{q}^{ }_{\,\bar\nu}) 
 \;.  \la{obs_2}
\ee
Finally, 
if we weigh the integrand, we may define
an energy density transfer rate from the QED plasma 
to an initially empty neutrino-antineutrino ensemble, 
\be
 \dot{e}^\rmi{gain}_{\nu + \bar\nu} \; \supset \; 
 \int_{\vec{k}^{ }_\nu,\vec{q}^{ }_{\,\bar\nu}} \hspace*{-1mm}
 (\, k^{ }_{\nu} +  q^{ }_{\bar\nu} \,) 
      \Psi(\vec{k}^{ }_{\nu},\vec{q}^{ }_{\,\bar\nu}) 
 \;. \la{obs_3}
\ee
We return to further 
concepts, including all factors from \eq\nr{b_with_f}, 
in \se\ref{se:energy}.

\vspace*{3mm}

Suppose now that we approximate \eq\nr{b_with_f} by considering 
an initial state in which the neutrino and antineutrino
are close to equilibrium. In the early universe, this is true
at the initial stages of the decoupling process. Then we may 
expand the neutrino and antineutrino 
phase space distributions to first order
around equilibrium, 
$
 f^{ }_{\vec{k}^{ }_{\nu}}
 \to 
 f^{ }_\rmii{F}(k^{ }_{\nu})
 + 
 \delta f^{ }_{\vec{k}^{ }_{\nu}} 
$
and 
$
 f^{ }_{\vec{q}^{ }_{\,\bar\nu}}
 \to 
 f^{ }_\rmii{F}(q^{ }_{\bar\nu})
 + 
 \delta f^{ }_{\vec{q}^{ }_{\,\bar\nu}} 
$.
The zeroth order term from this expansion
vanishes, because of detailed balance. 
The leading term (multiplied by $-1$)
comes from the first correction,
\ba
 &&
  f^{ }_{\vec{k}^{ }_\nu}
  f^{ }_{\vec{q}^{ }_{\,\bar\nu}}
  [1 - f^{ }_\rmii{F}({\epsilon}^{ }_{e})] 
  [1 - f^{ }_\rmii{F}({\epsilon}^{ }_{\bar e})] 
 - 
  f^{ }_\rmii{F}({\epsilon}^{ }_{e})
  f^{ }_\rmii{F}({\epsilon}^{ }_{\bar e})
  ( 1 - f^{ }_{\vec{k}^{ }_\nu} ) 
  ( 1 -   f^{ }_{\vec{q}^{ }_{\,\bar\nu}} ) 
 \nn[2mm]
 & \to &
  \delta f^{ }_{\vec{k}^{ }_\nu}
  \, \bigl\{ \, 
  f^{ }_\rmii{F}({q}^{ }_{\bar\nu} )
  [1 - f^{ }_\rmii{F}({\epsilon}^{ }_{e})] 
  [1 - f^{ }_\rmii{F}({\epsilon}^{ }_{\bar e})] 
 + 
  f^{ }_\rmii{F}({\epsilon}^{ }_{e})
  f^{ }_\rmii{F}({\epsilon}^{ }_{\bar e})
  [ 1 - f^{ }_\rmii{F}({q}^{ }_{\bar\nu} ) ] 
 \,\bigr\} 
 + (\vec{k}^{ }_\nu \leftrightarrow \vec{q}^{ }_{\,\bar\nu})
 + \rmO(\delta^2_{ })
 \hspace*{5mm}
 \nn[2mm]
 & = & 
  \delta f^{ }_{\vec{k}^{ }_\nu}
  \, 
  f^{ }_\rmii{F}({q}^{ }_{\bar\nu} )
  f^{ }_\rmii{F}({\epsilon}^{ }_{e})
  f^{ }_\rmii{F}({\epsilon}^{ }_{\bar e}) 
  \bigl[\,
   e^{( {\epsilon}^{ }_{e} + {\epsilon}^{ }_{\bar e})/T^{ }_\gamma }_{ }
  + 
   e^{q^{ }_{\bar\nu}/T^{ }_\gamma }_{ }
  \,\bigr]
  + (\vec{k}^{ }_\nu \leftrightarrow \vec{q}^{ }_{\,\bar\nu})
  + \rmO(\delta^2_{ })
  \nn[2mm]
 & = & 
  \delta f^{ }_{\vec{k}^{ }_\nu}
  \, 
  f^{-1}_\rmii{B}( {\epsilon}^{ }_{e} + {\epsilon}^{ }_{\bar e} )
  f^{ }_\rmii{F}({\epsilon}^{ }_{e})
  f^{ }_\rmii{F}({\epsilon}^{ }_{\bar e}) 
  \bigl[\,
    f^{ }_\rmii{F}({q}^{ }_{\bar\nu} )
  + 
    f^{ }_\rmii{B}( {\epsilon}^{ }_{e} + {\epsilon}^{ }_{\bar e} )
  \,\bigr]
  + (\vec{k}^{ }_\nu \leftrightarrow \vec{q}^{ }_{\,\bar\nu})
  + \rmO(\delta^2_{ })
  \nn[2mm]
 & = & 
  \delta f^{ }_{\vec{k}^{ }_\nu}
  \, 
  f^{-1}_\rmii{B}( p^{ }_0 )
  f^{ }_\rmii{F}({\epsilon}^{ }_{e})
  f^{ }_\rmii{F}({\epsilon}^{ }_{\bar e}) 
  \bigl[\,
    f^{ }_\rmii{F}( p^{ }_0 - k^{ }_{\nu} )
  + 
    f^{ }_\rmii{B}( p^{ }_0 )
  \,\bigr]^{ }_{p^{ }_0 \,\equiv\,
  {\epsilon}^{ }_{e} + {\epsilon}^{ }_{\bar e} }
  + (\vec{k}^{ }_\nu \leftrightarrow \vec{q}^{ }_{\,\bar\nu})
  + \rmO(\delta^2_{ })
 \;, \la{boltzmann_expanded}
\ea 
where $f^{ }_\rmii{B}$ denotes a Bose distribution. 
Inserting into \eq\nr{b_with_f}, we find
\ba
  \dot{f}^{ }_{\vec{k}^{ }_\nu}
 & = & 
  - \Gamma^{ }_{\vec{k}^{ }_\nu}
    \delta f^{ }_{\vec{k}^{ }_\nu}
  - 
  \int_{\vec{q}^{ }_{\,\bar\nu}}
    \Upsilon^{ }_{\vec{k}^{ }_\nu,\vec{q}^{ }_{\,\bar\nu}}
  \delta f^{ }_{\vec{q}^{ }_{\,\bar\nu}}
  + \rmO(\delta^2_{ })
 \;, \la{lin_b_1} \\[2mm] 
  \dot{f}^{ }_{\vec{q}^{ }_{\,\bar\nu}}
 & = & 
  - \Gamma^{ }_{\vec{q}^{ }_{\,\bar\nu}}
    \delta f^{ }_{\vec{q}^{ }_{\,\bar\nu}}
  - 
  \int_{\vec{k}^{ }_{\nu}}
    \Upsilon^{ }_{\vec{q}^{ }_{\,\bar\nu},\vec{k}^{ }_\nu}
  \delta f^{ }_{\vec{k}^{ }_{\nu}}
  + \rmO(\delta^2_{ })
 \;, \la{lin_b_2} \\[2mm] 
 \Gamma^{ }_{\vec{k}^{ }_\nu}
 & \supset &
 \int_{\vec{q}^{ }_{\,\bar\nu}}
      \,f^{-1}_\rmii{B}( p^{ }_0 )
      \,\bigl[\,
        f^{ }_\rmii{F}( p^{ }_0 - k^{ }_{\nu} )
      + 
        f^{ }_\rmii{B}( p^{ }_0 )
      \,\bigr]
      \Psi(\vec{k}^{ }_{\nu},\vec{q}^{ }_{\,\bar\nu}) 
 \bigr|^{ }_{
  p^{ }_0 \, = \,
  k^{ }_\nu + q^{ }_{\bar\nu}
  }
 \;,  \la{def_Gamma}  \\[3mm]
 \Upsilon^{ }_{\vec{k}^{ }_\nu,\vec{q}^{ }_{\,\bar\nu}}
 & \supset &
      \,f^{-1}_\rmii{B}( p^{ }_0 )
      \,\bigl[\,
        f^{ }_\rmii{F}( p^{ }_0 - q^{ }_{\bar\nu} )
      + 
        f^{ }_\rmii{B}( p^{ }_0 )
      \,\bigr]
      \Psi(\vec{k}^{ }_{\nu},\vec{q}^{ }_{\,\bar\nu}) 
 \bigr|^{ }_{
  p^{ }_0 \, = \,
  k^{ }_\nu + q^{ }_{\bar\nu}
  }
 \;. \la{def_Upsilon}
\ea
The coefficient $ \Gamma^{ }_{\vec{k}^{ }_\nu} $ is 
referred to as the neutrino 
{\em interaction rate}.
As is visible in \eq\nr{def_Gamma},  
it originates from the same $\Psi$ that determines
the energy transfer rate in \eq\nr{obs_3}.

%
\subsection{Scattering}

We now turn to scattering (``$t$-channel'') processes, 
originating from elastic interactions  
of a neutrino of momentum $\vec{k}^{ }_{\nu}$.\footnote{%
  Of course, there will be similar results for antineutrinos.
}
Proceeding by analogy with \eq\nr{def_Psi}, we should define 
a {\em transition rate}  
$\vec{q}^{ }_{\,\nu} \to \vec{k}^{ }_{\nu}$,  
assuming electrons to be in thermal equilibrium. 
We denote this by
\be
 \Theta(\vec{q}^{ }_{\,\nu} \to \vec{k}^{ }_{\nu})
 \;\equiv\;
 \sum_{\pm}
 \frac{1}{ 2 k^{ }_\nu }
  \int^{ }_{\vec{p}^{ }_{e_i},\vec{p}^{ }_{e_f}}
  \hspace*{-5mm}
  {\rm d}\Phi^{ }_{\vec{q}^{ }_{\,\nu} + \vec{p}^{ }_{e_i} \to\, 
                   \vec{k}^{ }_{\nu}  + \vec{p}^{ }_{e_f} } 
  \,
  f^{ }_\rmii{F} ( \epsilon^{ }_{e_i} )
  [1- f^{ }_\rmii{F} ( \epsilon^{ }_{e_f} )]
  \;
 {\textstyle\sum}|\mathcal{M}|^2_{ }
 \;, \la{def_Theta}  
\ee
where the sum $\sum_\pm$ refers to electrons and positrons, and 
$e^{ }_{i,f}$ to the initial and final-state energies, respectively. 
The corresponding part of the Boltzmann equation then reads
\be
 \dot{f}^{ }_{\vec{k}^{ }_\nu}
 \supset
  \int^{ }_{\vec{q}^{ }_{\,\nu}}
 \Bigl[
 \underbrace{
   \Theta( 
   \vec{q}^{ }_{\,\nu} \to 
   \vec{k}^{ }_{\nu}
   )
  f^{ }_{\vec{q}^{ }_{\,\nu} }
  ( 1 - f^{ }_{\vec{k}^{ }_{\nu}} )
 }_{\rm gain}
 - 
 \underbrace{
   \Theta( 
   \vec{k}^{ }_{\nu} \to 
   \vec{q}^{ }_{\,\nu}
   )
  f^{ }_{\vec{k}^{ }_{\nu}} 
  ( 1 - f^{ }_{\vec{q}^{ }_{\,\nu} })
 }_{\rm loss}
 \Bigr]
 \;. \la{boltzmann_t}
\ee
The gain and loss terms are related by
\be
   \Theta( 
   \vec{q}^{ }_{\,\nu} \to
   \vec{k}^{ }_{\nu}
   )
  =
  e^{(q^{ }_{\,\nu} - k^{ }_{\nu})/T^{ }_\gamma}_{ }
  \; \Theta( 
   \vec{k}^{ }_{\nu} \to 
   \vec{q}^{ }_{\,\nu}
   ) 
  \;, \la{balance_2}
\ee
which ensures detailed balance:
$
 \nF^{ }(q^{ }_\nu) [1 - \nF^{ }(k^{ }_\nu)]
 = 
 e^{(k^{ }_\nu - q^{ }_\nu)/T^{ }_\gamma}_{ } \,
 \nF^{ }(k^{ }_\nu) [1 - \nF^{ }(q^{ }_\nu)]
$.

\vspace*{3mm}

Even if the number of electrons or neutrinos does not
change in elastic scatterings, 
energy can be transferred.
It thus seems natural, by analogy with \eq\nr{obs_3}, to consider
\be
 \dot{e}^\rmi{gain}_{\nu\to\nu} \; \supset \; 
 \int_{\vec{k}^{ }_\nu,\vec{q}^{ }_{\,\nu}} \hspace*{-1mm}
 (\, k^{ }_{\nu} -  q^{ }_{\nu} \,) 
      \Theta(\vec{q}^{ }_{\,\nu} \to \vec{k}^{ }_{\nu}) 
 f^{ }_{\vec{q}^{ }_{\,\nu}}
 \;, \quad 
 \mbox{for}~f^{ }_{\vec{k}^{ }_\nu} \ll 1
 \;. 
\ee
Further variants, including the proper Pauli factors, 
will be defined in \se\ref{se:energy}.

\vspace*{3mm}

In order to obtain the contribution of 
elastic scatterings to the neutrino interaction rate, 
we return to the case whereby
neutrinos are close to equilibrium, and
expand their phase space distributions to first order
around equilibrium, 
$
 f^{ }_{\vec{k}^{ }_{\nu}}
 \to 
 f^{ }_\rmii{F}(k^{ }_{\nu})
 + 
 \delta f^{ }_{\vec{k}^{ }_{\nu}} 
$
and 
$
 f^{ }_{\vec{q}^{ }_{\,\nu}}
 \to 
 f^{ }_\rmii{F}(q^{ }_{\nu})
 + 
 \delta f^{ }_{\vec{q}^{ }_{\,\nu}} 
$ in \eq\nr{boltzmann_t}.
As in \eq\nr{boltzmann_expanded}, 
but skipping the intermediate steps, this yields
\ba
 &&
  f^{ }_{\vec{q}^{ }_{\,\nu}}
  (1-f^{ }_{\vec{k}^{ }_\nu})
  f^{ }_\rmii{F}({\epsilon}^{ }_{e_i})
  [1 - f^{ }_\rmii{F}({\epsilon}^{ }_{e_f})] 
 - 
  f^{ }_{\vec{k}^{ }_\nu} 
  ( 1 -   f^{ }_{\vec{q}^{ }_{\,\nu}} ) 
  f^{ }_\rmii{F}({\epsilon}^{ }_{e_f})
  [1-f^{ }_\rmii{F}({\epsilon}^{ }_{e_i})]
 \nn[2mm]
  & \to &
  - \, \delta f^{ }_{\vec{k}^{ }_\nu}
  \, 
  f^{-1}_\rmii{B}( p^{ }_{0} )
  f^{ }_\rmii{F}({\epsilon}^{ }_{e_i})
  [1-f^{ }_\rmii{F}({\epsilon}^{ }_{e_f}) ]
  \bigl[\,
    f^{ }_\rmii{F}( p^{ }_{0} - k^{ }_{\nu} )
  + 
    f^{ }_\rmii{B}( p^{ }_{0} )
  \,\bigr]^{ }_{p^{ }_0 \,\equiv\,
  {\epsilon}^{ }_{e_i} - {\epsilon}^{ }_{e_f} }
  \nn[2mm]
  & & 
  + \, \delta f^{ }_{\vec{q}^{ }_{\,\nu}}
  \, 
  f^{-1}_\rmii{B}( p^{ }_{0} )
  [1-f^{ }_\rmii{F}({\epsilon}^{ }_{e_i})]
  f^{ }_\rmii{F}({\epsilon}^{ }_{e_f}) 
  \bigl[\,
    f^{ }_\rmii{F}( p^{ }_{0} - q^{ }_{\nu} )
  + 
    f^{ }_\rmii{B}( p^{ }_{0} )
  \,\bigr]^{ }_{p^{ }_0 \,\equiv\,
  {\epsilon}^{ }_{e_f} - {\epsilon}^{ }_{e_i} }
  + \rmO(\delta^2_{ }) \;. \hspace*{6mm}
\ea
For the terms in the linearised Boltzmann equations,
\eqs\nr{lin_b_1} and \nr{lin_b_2},  
we get
\ba
 \Gamma^{ }_{\vec{k}^{ }_\nu}
 & \supset &
 \int_{\vec{q}^{ }_{\,\nu}}
      \,f^{-1}_\rmii{B}( p^{ }_0 )
      \,\bigl[\,
        f^{ }_\rmii{F}( p^{ }_0 - k^{ }_{\nu} )
      + 
        f^{ }_\rmii{B}( p^{ }_0 )
      \,\bigr]
      \Theta(\vec{q}^{ }_{\,\nu} \to \vec{k}^{ }_{\nu}) 
 \bigr|^{ }_{
  p^{ }_0 \, = \,
  k^{ }_\nu - q^{ }_{\nu}
  }
 \;,  \la{def_Gamma_t}  \\[3mm]
 \Upsilon^{ }_{\vec{k}^{ }_\nu,\vec{q}^{ }_{\,\nu}}
 & \supset &
    - \,f^{-1}_\rmii{B}( p^{ }_0 )
      \,\bigl[\,
        f^{ }_\rmii{F}( p^{ }_0 - q^{ }_{\nu} )
      + 
        f^{ }_\rmii{B}( p^{ }_0 )
      \,\bigr]
      \Theta(\vec{k}^{ }_{\nu} \to \vec{q}^{ }_{\,\nu}) 
 \bigr|^{ }_{
  p^{ }_0 \, = \,
  q^{ }_\nu - k^{ }_{\nu}
  }
 \;.
\ea

%
\section{Quantum-mechanical derivation of double-differential rates}
\la{se:qm}

Inspired by the intuitive picture from \se\ref{se:intuition}, our goal
now is to define double-differential neutrino-antineutrino production
and scattering rates
directly within quantum statistical physics, generalizing thereby on 
\eqs\nr{def_Psi} and \nr{def_Theta}, so that NLO QED corrections can be 
included. In order to go in this direction, we first collect
together some definitions.  

%
\subsection{Notation and conventions}
\la{ss:conventions}

We assume that neutrinos interact with a QED plasma
as dictated by a Fermi effective theory. After a Fierz transformation, 
the interaction Hamiltonian can be expressed as 
\be
 H^{ }_\iI(t) \; \supset \; \int_{\vec{x}}
 \bar{\nu}^{ }_{a} \gamma^{ }_\mu (1-\gamma^{ }_5) \nu^{ }_{a} 
 \,
 \underbrace{
 \frac{G^{ }_\rmii{F}}{2\sqrt{2}}
 \biggl\{ \,
    \, \bar{\ell}^{ }_e  \gamma^\mu _{ }
           \biggl[ 
                   2\delta^{ }_{a,e} - 1 + 4 \xW
                 +  \frac{ e^2_{ } C^{ }_a }{2} 
                 + (1 - 2\delta^{ }_{a,e} )\gamma^{ }_5
           \biggr] \ell^{ }_e 
 \biggr\}
 }_{\,\equiv\,\mathcal{O}^\mu_{ }(\X)}
 \;, \hspace*{5mm}
 \la{H_I_full}
\ee
where $\X \equiv (t,\vec{x})$, 
$\int_\vec{x} \equiv \int \! {\rm d}^3_{ }\vec{x}$,
$a$ is a flavour index, 
$G^{ }_\rmii{F}$ is the Fermi coupling, 
$\xW \equiv \sin^2_{ } \theta^{ }_\aW$ where 
$\theta^{ }_\aW$ is the Weinberg angle, 
$e^2_{ }\equiv 4\pi \alpha^{ }_\rmi{em}$ is the QED gauge coupling, 
and $C^{ }_a$ is a matching coefficient from the NLO construction of 
the Fermi effective theory
(cf.\ \eq\nr{C_a}). 

In addition to interacting with the QED plasma, the neutrinos also
interact with themselves and with other neutrinos. 
However, those 
interactions do not give rise to NLO QED corrections, and therefore
play no role in the current investigation. 

We do need to account for the interactions of electrons with each
other, through the QED Lagrangian. There is a class of processes, 
in which the electrons interact via soft $t$-channel 
photon exchange, in which the photon can be ``soft'' and 
needs to be Hard Thermal Loop (HTL)
resummed~\cite{rate}. However, the electrons participating in 
this reaction are ``hard'', with momenta $p \sim \pi T$, so we 
do not need to employ HTL vertices~\cite{htl6}. Therefore it is 
sufficient to supplement \eq\nr{H_I_full} by the vertex
\be
 \mathcal{L}^{ }_\rmii{QED} \;\supset\; 
 e\,
 \bar{\ell}^{ }_e \gamma^\mu_{ } \ell^{ }_e A^{ }_\mu
 \;, \la{L_QED}
\ee
use a tree-level propagator for electrons, and a HTL-resummed
propagator for photons. The latter will be denoted by 
$
 \Delta^{-1*}_{\P;\mu\nu}
$.

The neutrino fields (in the interaction picture) 
can be expressed in a mode expansion, 
\ba
 \bar\nu^{ }_a(\X) & = &  
 \int_\vec{k} \frac{1}{\sqrt{2k}}
 \biggl(
   \bar{u}^{ }_{\vec{k}a}\,\hat{w}^{\dagger}_{\vec{k}a}\, e^{+i \K\cdot\X }_{ }
 + 
   \bar{v}^{ }_{\vec{k}a}\,\hat{x}^{ }_{\vec{k}a}\, e^{- i \K\cdot\X}_{ }   
 \biggr)
 \;, \la{bar_nu_a}
 \\[2mm]
 \nu^{ }_a(\X) & = &  
 \int_\vec{q} \frac{1}{\sqrt{2q}}
 \biggl(
   u^{ }_{\vec{q}a}\,\hat{w}^{ }_{\vec{q}a}\, e^{- i \Q\cdot\X }_{ }
 + 
   v^{ }_{\vec{q}a}\,\hat{x}^\dagger_{\vec{q}a}\, e^{+ i \Q\cdot\X}_{ }   
 \biggr)
 \;, \la{nu_a}
\ea
where 
$
 \K\cdot\X = k\, t - \vec{k}\cdot\vec{x}
$.
The creation and annihilation operators
of neutrinos ($\hat{w}^\dagger_{\vec{k}a}$, 
$\hat{w}^{ }_{\vec{k}a}$) 
and antineutrinos ($\hat{x}^\dagger_{\vec{q}a}$, 
$\hat{x}^{ }_{\vec{q}a}$) 
are normalized as 
\be
 \{ \hat{w}^{ }_{\vec{k}a},\hat{w}^\dagger_{\vec{q}b} \}
 \; = \; 
 \{ \hat{x}^{ }_{\vec{k}a},\hat{x}^\dagger_{\vec{q}b} \}
 \; = \; 
 (2\pi)^3_{ }\,\delta^{(3)}_{ }(\vec{k-q})\,\delta^{ }_{ab}
 \;. \la{commutator}
\ee
With these normalizations, 
a neutrino density matrix, with its diagonal components corresponding
to the classical phase space distribution, reads
\be
 \hat{\rho}^{ }_{\vec{k}ab}
 \; \equiv \; 
 \frac{\hat{w}^\dagger_{\vec{k}a} \hat{w}^{ }_{\vec{k}b}}{V}
 \;, \quad
  \bigl\langle\,
   \hat{\rho}^{ }_{\vec{k}ab}   
  \,\bigr\rangle
 \; = \; 
 f^{ }_{\vec{k} a} \,\delta^{ }_{ab}
 \;, \la{def_f_qm}
\ee
where $V$ denotes a volume and 
$\langle ... \rangle$ an ensemble average. 
The volume appears as a ``regulator''  in \eq\nr{def_f_qm}
(also in \eq\nr{def_dot_f_qm}), as we have set the momenta
to coincide, and \eq\nr{commutator} indicates
that this limit is singular, however it 
drops out later on (see below). 

%
\subsection{Production and annihilation rates of a neutrino-antineutrino pair}
\la{ss:qm_prod}

Let us now define an initial and a final state of the type
\be
 |I\,\rangle \equiv |\ini\rangle \otimes 
 | 0 \rangle 
 \;, \quad
 |F\,\rangle \equiv |\fin\rangle \otimes 
 | \vec{k}^{ }_\nu,\vec{q}^{ }_{\,\bar\nu} \rangle
 \;, \la{states_prod}
\ee
where $|\ini\rangle$ and $|\fin\rangle$ are states in the QED Fock space; 
$
 | \vec{k}^{ }_\nu,\vec{q}^{ }_{\,\bar\nu} \rangle
 = 
 \hat{w}^\dagger_{ \vec{k}^{ }_\nu }
 \hat{x}^\dagger_{ \vec{q}^{ }_{\,\bar\nu } }
 | 0 \rangle
$; 
$
 | 0 \rangle
$
refers to the neutrino vacuum; 
and flavour indices 
have been omitted
for simplicity. 
To first order in the interaction Hamiltonian, 
the transition matrix element reads
\be
 T^{ }_\rmi{$F$$I$} =
 \langle F\, | \int_0^{t} \! {\rm d}t' \, \hat H_\iI^{ }(t') \, 
 | I\,\rangle
 \;. \la{transition_matrix}
\ee
A quantum-mechanical version of the 
fully inclusive production rate from \eqs\nr{def_Psi}
is then defined as 
\be
 \Psi( \vec{k}^{ }_{\nu},\vec{q}^{ }_{\,\bar\nu} )
 \; \equiv \; 
 \lim_{t,V\to \infty} 
 \sum_{\sini,\,\sfin}
 \frac{e^{-E^{ }_{\sini}/T^{ }_\gamma}_{ }}{\mathcal{Z}^{ }_\rmiii{QED}}
 \frac{| T^{ }_\rmii{$F$$I$}|^2_{ }}{t\, V}
 \;, \la{def_dot_f_qm}
\ee
where $E^{ }_{\sini}$ are the eigenvalues of the QED Hamiltonian, 
and 
$
 \mathcal{Z}^{ }_\rmii{QED} 
 \equiv 
 \sum_{\sini} e^{-E^{ }_{\sini}/T^{ }_\gamma}_{ }
$.

Inserting the field operators from \eqs\nr{bar_nu_a} and \nr{nu_a}
into the Hamiltonian from \eq\nr{H_I_full}, we obtain
\be
 T^{ }_\rmi{$F$$I$} = 
 \frac{\bar{u}^{ }_{ \vec{k}^{ }_\nu } \gamma^{ }_\mu (1 - \gamma^{ }_5) 
           {v}^{ }_{ \vec{q}^{ }_{\,\bar\nu} } }
      { \sqrt{ (2 k^{ }_\nu)(2 q^{ }_{\bar\nu}) } }
 \,
 \int_{\X'}
 \,
 \langle \fin | \mathcal{O}^\mu_{ }(\X') | \ini \rangle
 \; 
 e^{i (\K^{ }_\nu + \Q^{ }_{\bar\nu})\cdot \X' }_{ }
 \;. 
\ee
Once we employ this in \eq\nr{def_dot_f_qm}, and make use of spacetime
translational invariance, one integral can be factored out, 
and cancelled against
the denominator. For the neutrino spinors, we can sum over would-be 
spins (even though only one helicity state is physical),
obtaining  
\ba
 {\textstyle\sum}\;
 \bar{u}^{ }_{\vec{k}^{ }_\nu} 
 \gamma^{ }_{\mu} (1 - \gamma^{ }_5)
 {v}^{ }_{\vec{q}^{ }_{\,\bar\nu}} 
 \bar{v}^{ }_{\vec{q}^{ }_{\,\bar\nu}} 
 \gamma^{ }_{\bar\mu} (1 - \gamma^{ }_5)
 u^{ }_{\vec{k}^{ }_\nu} \, 
 & = & 
 \tr \bigl[\, 
 {\bsl\K}^{ }_{ }
 \gamma^{ }_{\mu} (1 - \gamma^{ }_5)\,
 {\bsl\Q}^{ }_{ } 
 \gamma^{ }_{\bar\mu} (1 - \gamma^{ }_5)
 \,\bigr]
 \nn[2mm]
 & \supset & 
 8 \bigl(\,
 \underbrace{
    \K^{ }_\mu  \Q^{ }_{\bar\mu} 
  + \K^{ }_{\bar\mu} \Q^{ }_{\mu} 
  - \eta^{ }_{\mu\bar\mu}\, 
    \K\cdot\Q}_{
    \equiv\, L_{\mu\bar\mu}^{ }(\K,\Q)
  }
 \,\bigr)
 \;, \la{def_L}
\ea
where $\eta \equiv\,$diag($+$$-$$-$$-$).
In \eq\nr{def_L} we anticipated that the result will be contracted with
a tensor symmetric in $\mu\leftrightarrow\bar\mu$, 
namely the vector channel spectral function 
$\im V^{\mu\bar\mu}_{ }$,
so that
an antisymmetric part originating from $\gamma^{ }_5$ can be omitted. 
In addition we had dropped the neutrino and antineutrino
labels from $\K$ and $\Q$, in order to avoid double subscripts, 
and from now on we will do the same with $k$ and $q$.
For \eq\nr{def_dot_f_qm}, this yields
\be
 \Psi ( \vec{k}^{ }_{\nu},\vec{q}^{ }_{\,\bar\nu} )
 \; = \; 
 \frac{2
 \, L_{\mu\bar\mu}(\K,\Q)
        }{k\, q}
 \underbrace{ 
 \int_{\X}
 \bigl\langle\,
   \mathcal{O}^{\bar\mu}_{ }(0) \, 
   \mathcal{O}^{\mu}_{ }(\X) 
 \,\bigr\rangle^{ }_{\T_{\!\gamma}}
 \, e^{i(\K+\Q)\cdot\X }_{ }
 }_{ 
 \Pi^{\mu\bar\mu,<}_{\K+\Q}
 \;=\; 
 2 f^{ }_\rmiii{B}(k^{ }_{ } + q^{ }_{ })\,
 \rho^{ \mu\bar\mu }_{\K + \Q}
 }
 \;. \la{wight_1}
\ee
In the last step, we denoted by 
$
 \langle...\rangle^{ }_{\T_{\!\gamma}}
 \equiv 
 \sum_{\sini} e^{-E^{ }_{\sini}/T^{ }_\gamma}_{ } \langle
 \ini |... | \ini\rangle / 
 \mathcal{Z}^{ }_\rmii{QED} 
$ 
a thermal
expectation value in the QED ensemble; 
identified the correlator as a Wightman function $\Pi^{<}_{ }$;
and expressed it in terms of 
the corresponding spectral function, $\rho$,
via a text-book relation. 

In order to make the expression more concrete, we insert
the operator $\mathcal{O}^\mu_{ }$ from \eq\nr{H_I_full}. 
When we evaluate the corresponding spectral function, 
disconnected contributions, involving two insertions
of the vertex in \eq\nr{L_QED}, need to be included. 
Then, denoting the connected part of the vector
channel correlator by $V^{\mu\bar\mu}_{ }$, 
we get\hspace*{0.4mm}\footnote{%
 We have also put the electron  
 mass to zero here, which would not be necessary.
 In appendix~\ref{app:B}, we determine the effects from
 $m^{ }_e/T^{ }_\gamma > 0$
 at leading order in QED.  \la{fn:me}
 }
\ba
 \Psi( \vec{k}^{ }_{\nu},\vec{q}^{ }_{\,\bar\nu} )
 & = & 
 \frac{G^2_\rmiii{F}
 \, L_{\mu\bar\mu}(\K,\Q)
 }{2\,k\, q}
 \, f^{ }_\rmii{B}(k + q) 
 \nn[2mm]
 & \times & 
 \biggl\{\,
  \biggl[
    \biggl( 
      2\delta^{ }_{a,e} - 1 + 4 \xW
                 +  \frac{ e^2_{ } C^{ }_a }{2}
    \biggr)^2_{ } + 1 
  \biggr]
  \,   \im V^{\mu\bar\mu}_{\K+\Q}
 \nn[2mm]
 & - &
 e^2_{ }
 \bigl( 
      2\delta^{ }_{a,e} - 1 + 4 \xW
 \bigr)^2_{ }
 \im\Bigl[\,
  V^{\mu\alpha}_{\K+\Q}
  \, \Delta^{-1*}_{\K+\Q;\alpha\beta}
  \, V^{\beta\bar\mu}_{\K+\Q}
  \,\Bigr] 
 \,\biggr\}
 + \rmO(e^4_{ })
  \;. \la{res_1}
\ea
For brevity we have expressed the counterterm contribution inside
a square, but the expression is complete only up to and including
the order $\rmO(e^2_{ })$. After some further rewriting, this 
expression turns into one of our main results, given by 
\eqs\nr{Psi_NLO} and \nr{calF} below.

\vspace*{3mm}

In order to define a double-differential annihilation rate, 
we exchange the roles of the neutrino states in \eq\nr{states_prod}, 
{\it viz.}\
\be
 |I\,\rangle \equiv |\ini\rangle \otimes 
 | \vec{k}^{ }_\nu,\vec{q}^{ }_{\,\bar\nu} \rangle
 \;, \quad
 |F\,\rangle \equiv |\fin\rangle \otimes 
 | 0 \rangle 
 \;. \la{states_anni}
\ee
The computation proceeds as before, however the Wightman
correlator in \eq\nr{wight_1} is replaced with 
\be
 \Pi^{\mu\bar\mu,>}_{\K+\Q}
 \;=\; 
 2
 \, [1 + f^{ }_\rmiii{B}(k^{ }_{ } + q^{ }_{ }) ]\,
 \, \rho^{ \mu\bar\mu }_{\K + \Q}
 \;=\; 
 2
 \, e^{(k^{ }_{ } + q^{ }_{ } )/T^{ }_\gamma}_{ }
 \, f^{ }_\rmiii{B}(k^{ }_{ } + q^{ }_{ })
 \, \rho^{ \mu\bar\mu }_{\K + \Q}
 \;.
\ee
Thereby we obtain an annihilation rate, 
$\widetilde\Psi( \vec{k}^{ }_\nu,\vec{q}^{ }_{\,\bar\nu} )$, 
related to the production
rate by precisely the same detailed-balance relation 
as in \eq\nr{balance_1}.

%
\subsection{Scattering rates of a neutrino and an antineutrino}
\la{ss:qm_scat}

To promote the $t$-channel rate from \eq\nr{def_Theta} 
into a quantum-mechanical version 
of the same quantity, 
we follow closely the steps of \se\ref{ss:qm_prod}. 
We focus on the gain term in \eq\nr{boltzmann_t}, 
whereby the initial and final states are 
\be
 |I\,\rangle \equiv |\ini\rangle \otimes 
 | \vec{q}^{ }_{\,\nu} \rangle 
 \;, \quad
 |F\,\rangle \equiv |\fin\rangle \otimes 
 | \vec{k}^{ }_{\nu} \rangle
 \;, \la{states_t}
\ee
where 
$
 | \vec{k}^{ }_\nu \rangle
 = 
 \hat{w}^\dagger_{ \vec{k}^{ }_\nu }
 | 0 \rangle
$ 
and 
$
 | \vec{q}^{ }_{\,\nu} \rangle
 = 
 \hat{w}^\dagger_{ \vec{q}^{ }_{\,\nu} }
 | 0 \rangle
$. 
Again the task is to compute the transition matrix
element in \eq\nr{transition_matrix}, 
now with the states in \eq\nr{states_t}.
In this case one obtains
\be
 T^{ }_\rmi{$F$$I$} = 
 \frac{\bar{u}^{ }_{ \vec{k}^{ }_\nu } \gamma^{ }_\mu (1 - \gamma^{ }_5) 
           {u}^{ }_{ \vec{q}^{ }_{\,\nu} } }
      { \sqrt{ (2 k^{ }_\nu)(2 q^{ }_{\nu}) } }
 \,
 \int_{\X'}
 \,
 \langle \fin | \mathcal{O}^\mu_{ }(\X') | \ini \rangle
 \; 
 e^{i (\K^{ }_\nu - \Q^{ }_{\nu})\cdot \X' }_{ }
 \;. 
\ee
As in the previous section, we assume translational invariance 
and drop the antisymmetric~$\gamma_5$ tensors from the neutrino spin-sum, 
because their contribution drops out after a contraction with a symmetric
tensor. 
Using the same notation
as in \eq\nr{wight_1}, we can express 
the rate from \eq\nr{def_dot_f_qm}
in a similar form as \eq\nr{wight_1}, {\it viz.}\ 
\be
 \Theta( \vec{q}^{ }_{\,\nu} \to \vec{k}^{ }_{\nu} )
 \; = \; 
 \frac{2
 \, L_{\mu\bar\mu}(\K,\Q)
        }{k\, q}
 \underbrace{ 
 \int_{\X}
 \bigl\langle\,
   \mathcal{O}^{\bar\mu}_{ }(0) \, 
   \mathcal{O}^{\mu}_{ }(\X) 
 \,\bigr\rangle^{ }_{\T_{\!\gamma}}
 \, e^{i(\K-\Q)\cdot\X }_{ }
 }_{ 
 \Pi^{\mu\bar\mu,<}_{\K-\Q}
 \;=\; 
 2 f^{ }_\rmiii{B}(k^{ }_{ } - q^{ }_{ })\,
 \rho^{ \mu\bar\mu }_{\K - \Q}
 }
%
 \;. \la{res_t_1}
\ee
Exchanging the final and initial neutrino states, 
$\vec{q}^{ }_{\,\nu} \leftrightarrow\vec{k}^{ }_{\nu}\,$, 
and making use of the relations
\ba
 \rho^{ \mu\bar\mu }_{\Q - \K} 
 & = &
 - \rho^{ \mu\bar\mu }_{\K - \Q} 
 \;, \nn[2mm]
 f^{ }_\rmiii{B}(q^{ }_{ } - k^{ }_{ })
 & = &
 - e^{(k-q)/T^{ }_\gamma} \,
 f^{ }_\rmiii{B}(k^{ }_{ } - q^{ }_{ })
 \;,
\ea
leads to the same detailed balance relation as in \eq\nr{balance_2}, namely
\be
 \Theta( \vec{k}^{ }_{\nu}\to\vec{q}^{ }_{\,\nu} )
 \; = \;
 e^{(k-q)/T^{ }_\gamma}\,
 \Theta( \vec{q}^{ }_{\,\nu}\to\vec{k}^{ }_{\nu} )
 \;.
\ee

%
\subsection{Ideal choice of kinematic variables}

In order to make practical use of the double-differential rates
obtained in \eqs\nr{res_1} and \nr{res_t_1}, 
it is helpful to change variables. 
For the $s$-channel contribution,
we define the 4-momentum of the neutrino-antineutrino
pair as 
\be
 \P \; \equiv \; \K + \Q
 \;. 
\ee
Consequently, the antineutrino 4-momentum is 
$\Q = \P - \K$. From the identity
$
 \Q^2_{ } = (\P - \K)^2_{ } = 0
$, 
useful relations follow, in particular
\be
 \vec{k}\cdot\vec{p} = k p^{ }_0  - \frac{\P^2_{ }}{2}
 \;, \quad
 k^2_{ } - \frac{( \vec{k}\cdot\vec{p} )^2_{ }}{p^2_{ }}
 = 
 \frac{\P^2_{ }}{4}
 \biggl[
  1 - \biggl( \frac{2 k - p^{ }_0}{p} \biggr)^2_{ } 
 \biggr]
 \;. \la{relations}
\ee 

We now decompose the vector channel (i.e.\ photon) spectral function 
$
 \im V^{\mu\bar\mu}_{\P}
$
into its two independent parts, $\rho^{ }_\rmii{T}$, $\rho^{ }_\rmii{L}$, as
\be
 \im V^{\mu\bar\mu}_{\P} = 
 - {\eta^{\mu}_{ }}^{ }_i {\eta^{\,\bar\mu}_{ }}^{ }_j
 \biggl( \delta^{ }_{ij} - \frac{p^{ }_i\, p^{ }_j}{p^2_{ }} \biggr)
 \bigl(\, \rho^{ }_\rmii{T} - \rho^{ }_\rmii{L}\, \bigr)
 + 
 \biggl(
   \eta^{\mu\bar\mu}_{ } - \frac{\P^{\mu}_{ }\P^{\bar\mu}_{ }}{\P^2_{ }} 
 \biggr)
 \, \rho^{ }_\rmii{L}
 \;. \la{decomposition}
\ee
For reference, the LO expressions of 
$\rho^{ }_\rmii{T}$ and $\rho^{ }_\rmii{L}$
are given in \eqs\nr{rho_LO_T} and \nr{rho_LO_L}
in terms of integral representations, and in 
\eqs\nr{basis_trafo}--\nr{rho00} in closed form. 
Making use of \eq\nr{relations}, we then get
\ba
  L_{\mu\bar\mu}^{ }(\K, \P - \K ) 
  \im V^{\mu\bar\mu}_{\P} \,
 & = & 
 -\frac{\P^2_{ }}{2}
  \, 
  \biggl[\, 
    \rho^{ }_\rmii{T} + \rho^{ }_\rmii{L} 
  + 
   \biggl( \frac{2 k - p^{ }_0}{p} \biggr)^2_{ }
   \bigl(\, \rho^{ }_\rmii{T} - \rho^{ }_\rmii{L} \,\bigr)
  \,\biggr]
 \;.  \la{s_contraction}
\ea
Thereby, \eq\nr{res_1} can be expressed as 
\be
 \Psi( \vec{k}^{ }_{\nu},\vec{q}^{ }_{\,\bar\nu} )
 \; = \; 
 -\, \frac{
     G^2_\rmiii{F}
 \, f^{ }_\rmii{B}(k + q) 
   }{4\,k\, q} 
 \, \F^{ }(k;k+q,|\vec{k}^{ }_\nu + \vec{q}^{ }_{\,\bar\nu} |)
  + \rmO(e^4_{ })
 \;, \la{Psi_NLO}
\ee
where
\ba
 \F(k;p^{ }_0,p)  
 & \equiv & 
 \P^2_{ } \, \biggl\{\, 
  \biggl[
    \biggl( 
      2\delta^{ }_{a,e} - 1 + 4 \xW
                 +  \frac{ e^2_{ } C^{ }_a }{2}
    \biggr)^2_{ } + 1 
  \biggr]
 \la{calF}
 \\[2mm]
 & \times &  
  \biggl[\, 
    \rho^\rmii{NLO}_\rmii{T} + \rho^\rmii{NLO}_\rmii{L} 
  + 
   \biggl( \frac{2 k - p^{ }_0}{p} \biggr)^2_{ }
   \bigl(\, \rho^\rmii{NLO}_\rmii{T} - \rho^\rmii{NLO}_\rmii{L} \,\bigr)
  \,\biggr]
 \nn[2mm]
 & + & 
 2 e^2_{ } \bigl( 2\delta^{ }_{a,e} - 1 + 4 \xW \bigr)^2_{ }
 \nn[2mm]
 & \times & 
 \biggl[
    \rho^\rmii{LO}_\rmii{T} \chi^\rmii{LO}_\rmii{T} \mathcal{R}^*_\rmii{T}
  + \rho^\rmii{LO}_\rmii{L} \chi^\rmii{LO}_\rmii{L} \mathcal{R}^*_\rmii{L}
  + 
   \biggl( \frac{2k - p^{ }_0}{p} \biggr)^2_{ } 
   \bigl(
    \rho^\rmii{LO}_\rmii{T} \chi^\rmii{LO}_\rmii{T} \mathcal{R}^*_\rmii{T}
  - \rho^\rmii{LO}_\rmii{L} \chi^\rmii{LO}_\rmii{L} \mathcal{R}^*_\rmii{L}
   \bigr)  
 \biggr]
 \,\biggr\}
 \;. \nonumber 
\ea
Here $\chi^{ }_\rmii{T,L}$ originate from 
$\re V$ and 
$\mathcal{R}^{*}_\rmii{T,L}$ from $\Delta^{-1*}_{ }$
(cf.\ appendix~\ref{app:A}).

For the $t$-channel contribution, 
we instead change variables to the difference 
between neutrino 4-momenta,  
\be
 \P \; \equiv \; \K - \Q
 \; ,
\ee
which gives the same relations in \eq\nr{relations}. 
The contraction with the leptonic tensor differs 
from \eq\nr{s_contraction} by an overall minus sign, 
\ba
  L_{\mu\bar\mu}^{ }(\K,\K - \P)\, 
  \im V^{\mu\bar\mu}_{\P} \,
 & = & 
 \frac{\P^2_{ }}{2}
  \, 
  \biggl[\, 
    \rho^{ }_\rmii{T} + \rho^{ }_\rmii{L} 
  + 
   \biggl( \frac{2 k - p^{ }_0}{p} \biggr)^2_{ }
   \bigl(\, \rho^{ }_\rmii{T} - \rho^{ }_\rmii{L} \,\bigr)
  \,\biggr]
 \;. 
\ea
All in all this leads to a result similar to \eq\nr{Psi_NLO},
\be
 \Theta( \vec{q}^{ }_{\,\nu}\to\vec{k}^{ }_{\nu} )
 \; = \;
 +\, \frac{
     G^2_\rmiii{F}
 \, f^{ }_\rmii{B}(k-q) 
   }{4\,k\,q } 
  \, \F^{ }(k;k-q,|\vec{k}^{ }_\nu - \vec{q}^{ }_{\,\nu} |)
  + \rmO(e^4_{ })
  \;. \la{Theta_NLO}
\ee

%
\subsection{From double-differential rates to the neutrino interaction rate}
\la{ss:Gamma_nu}

In order to crosscheck that 
\eqs\nr{Psi_NLO} and \nr{Theta_NLO} are sensible, let us extract
the corresponding neutrino interaction rate. This follows from 
\eqs\nr{def_Gamma} 
and \nr{def_Gamma_t}, 
where $f^{-1}_\rmii{B}(p_0^{ })$ cancels 
against the corresponding factor in \eqs\nr{Psi_NLO}
and \nr{Theta_NLO}, respectively. 
The remaining 
phase space distributions 
in \eqs\nr{def_Gamma} 
and \nr{def_Gamma_t}
can be expressed as 
\be
 f^{ }_\rmii{F}( p^{ }_0 - k^{ }_{ } )
      + 
        f^{ }_\rmii{B}( p^{ }_0 )
 = 
 1 - f^{ }_\rmii{F}(  k^{ }_{ } - p^{ }_0  )
      + 
        f^{ }_\rmii{B}( p^{ }_0 )
 \;, \la{weight}
\ee 
setting them in the same form as in ref.~\cite{rate}. 
Furthermore, the $s$-channel measure for $\vec{q}$ reads
\ba
 \int_\vec{q}^{(s)} 
 & \equiv & 
 \int\! \frac{{\rm d}^3_{ }\vec{q}}{(2\pi)^3_{ }}
 \int\! {\rm d}^3_{ }\vec{p} \int \! {\rm d}p^{ }_0
 \, \delta^{(3)}_{ }(\vec{k}+\vec{q} - \vec{p})
 \, \delta(k + q - p^{ }_0)
 \nn[2mm]
 & = & 
 \int\! \frac{{\rm d}^3_{ }\vec{p}}{(2\pi)^3_{ }}
 \int\! {\rm d}p^{ }_0
 \, \delta(k - p^{ }_0 + |\vec{p} - \vec{k}|)
 \nn[2mm]
 & = & 
 \frac{1}{4\pi^2_{ }}
 \int_{-1}^{+1} \! {\rm d}z 
 \int_0^\infty \! {\rm d}p\, p^2_{ }
 \int_k^\infty \! {\rm d}p^{ }_0 \, 
 \delta\bigl(\, 
  k - p^{ }_0 + \sqrt{p^2_{ } + k^2_{ }- 2 p\, k\, z}
 \,\bigr)
 \nn[2mm]
 & = & 
 \frac{1}{2\pi^2_{ }k}
 \int_0^k \! {\rm d}p^{ }_{-}
 \int_k^\infty \! {\rm d}p^{ }_{+} \, p (p^{ }_0 - k)
 \biggr|^{ }_{p^{ }_\pm \, \equiv \, 
 \frac{p^{ }_\rmiii{0} \pm \, p}{2}}
 \;. \la{measure_s}
\ea
Integrating over \eq\nr{Psi_NLO} with this measure, 
the relevant $s$-channel parts of \eq(3.3) of 
ref.~\cite{rate} are reproduced. 
For the $t$-channel, we are faced instead with the measure
\ba
 \int_\vec{q}^{(t)} 
 & \equiv & 
 \int\! \frac{{\rm d}^3_{ }\vec{q}}{(2\pi)^3_{ }}
 \int\! {\rm d}^3_{ }\vec{p} \int \! {\rm d}p^{ }_0
 \, \delta^{(3)}_{ }(\vec{k}-\vec{q} - \vec{p})
 \, \delta(k - q - p^{ }_0)
 \nn[2mm]
 & = & 
 \int\! \frac{{\rm d}^3_{ }\vec{p}}{(2\pi)^3_{ }}
 \int\! {\rm d}p^{ }_0
 \, \delta(k - p^{ }_0 - |\vec{k} - \vec{p}|)
 \nn[2mm]
 & = & 
 \frac{1}{4\pi^2_{ }}
 \int_{-1}^{+1} \! {\rm d}z 
 \int_0^\infty \! {\rm d}p\, p^2_{ }
 \int_{-\infty}^k \! {\rm d}p^{ }_0 \, 
 \delta\bigl(\, 
  k - p^{ }_0 - \sqrt{k^2_{ } + p^2_{ }- 2 k\, p\, z}
 \,\bigr)
 \nn[2mm]
 & = & 
 \frac{1}{2\pi^2_{ }k}
 \int_{-\infty}^0 \! {\rm d}p^{ }_{-}
 \int_0^k \! {\rm d}p^{ }_{+} \, p (k-p^{ }_0)
 \biggr|^{ }_{p^{ }_\pm \, \equiv \, \frac{p^{ }_\rmiii{0} \pm \, p}{2}}
 \;. \la{measure_t}
\ea
Integrating over \eq\nr{Theta_NLO} with this measure,
including the factor from \eq\nr{weight}, 
we recover the relevant $t$-channel parts of \eq(3.3) of 
ref.~\cite{rate}. 

%
\section{Energy transfer rates}
\la{se:energy}

%
\subsection{Basic definitions}

Having validated our NLO results against the known neutrino 
interaction rate (cf.\ \se\ref{ss:Gamma_nu}), 
we now move on to the energy transfer rates. 
The energy transfer rates (into the neutrino ensemble) 
can be written as
\be
 Q
 \; \equiv \;
   \dot{e}^\rmi{gain}_{\nu + \bar\nu} 
 - \dot{e}^\rmi{loss}_{\nu + \bar\nu} 
 + \dot{e}^\rmi{scat}_{\nu \to \nu} 
 + \dot{e}^\rmi{scat}_{\bar\nu \to \bar\nu}  \, ,
 \la{Q_def}
\ee
where 
\ba
 \dot{e}^\rmi{gain}_{\nu + \bar\nu} 
 & \equiv &
 \int_{\vec{k}^{ }_\nu,\vec{q}^{ }_{\,\bar\nu}}^{(s)} \hspace*{-1mm}
 \bigl(\, k^{ }_{\nu} +  q^{ }_{\bar\nu} \,\bigr) \;
      \Psi(\vec{k}^{ }_{\nu},\vec{q}^{ }_{\,\bar\nu}) \;
  (1 - f^{ }_{\vec{k}^{ }_\nu}) 
  (1 - f^{ }_{\vec{q}^{ }_{\,\bar\nu} }) 
  \;, \la{edot_gain}
  \\[2mm]
 \dot{e}^\rmi{loss}_{\nu + \bar\nu} 
 & \equiv &
 \int_{\vec{k}^{ }_\nu,\vec{q}^{ }_{\,\bar\nu}}^{(s)} \hspace*{-1mm}
 \bigl(\, k^{ }_{\nu} +  q^{ }_{\bar\nu} \,\bigr) \;
      {\widetilde\Psi}(\vec{k}^{ }_{\nu},\vec{q}^{ }_{\,\bar\nu})  \;
  f^{ }_{\vec{k}^{ }_\nu}
  f^{ }_{\vec{q}^{ }_{\,\bar\nu} }
  \;, \la{edot_loss}
  \\[2mm]
 \dot{e}^\rmi{scat}_{\nu \to\nu} 
 & \equiv &
 \int_{\vec{k}^{ }_\nu,\vec{q}^{ }_{\,\nu}}^{(t)} \hspace*{-1mm}
 \bigl(\, k^{ }_{\nu} -  q^{ }_{\nu} \,\bigr) \;
      {\Theta}(\vec{q}^{ }_{\,\nu} \to \vec{k}^{ }_{\nu})  \;
  f^{ }_{\vec{q}^{ }_{\,\nu}}
  (1-f^{ }_{\vec{k}^{ }_{\nu} })
  \;.  \la{edot_scat}
\ea
To obtain the scattering contribution in \eq\nr{edot_scat}, 
we can envisage weighting \eq\nr{b_with_f} with~$k^{ }_\nu$, 
integrating over $\vec{k}^{ }_\nu$, and 
exchanging $\vec{k}^{ }_{\nu} \leftrightarrow \vec{q}^{ }_{\nu}$
in the loss term.  

Our goal in the next sections 
is to evaluate \eqs\nr{edot_gain}--\nr{edot_scat} 
in three different approximations: 
at full LO (cf.\ \se\ref{ss:edot_lo}); 
at approximate LO, based on Maxwell-Boltzmann statistics but
improved with correction factors (cf.\ \se\ref{ss:edot_mb}); 
and at full NLO (cf.\ \se\ref{ss:edot_nlo}). 

%
\subsection{Full leading order}
\la{ss:edot_lo}

The LO expressions for the double-differential rates appearing in 
\eqs\nr{edot_gain}--\nr{edot_scat} read, 
from \eqs\nr{Psi_NLO} and \nr{Theta_NLO}, 
\ba
 \Psi( \vec{k}^{ }_{\nu},\vec{q}^{ }_{\,\bar\nu} )
 & = & 
 -\, \frac{
     G^2_\rmiii{F}
 \, f^{ }_\rmii{B}(p_0^{ })
   }{4\,k^{ }_{\nu}\, q^{ }_{\bar\nu}}
  \Bigl[
    \bigl( 
      2\delta^{ }_{a,e} - 1 + 4 \xW
    \bigr)^2_{ } + 1 
  \Bigr]
 \nn[2mm]
 & \times & 
   \P^2_{ } \,  
  \biggl[\, 
    \rho^\rmii{LO}_\rmii{T} + \rho^\rmii{LO}_\rmii{L} 
  + 
   \biggl( \frac{2 k - p^{ }_0}{p} \biggr)^2_{ }
   \bigl(\, \rho^\rmii{LO}_\rmii{T} - \rho^\rmii{LO}_\rmii{L} \,\bigr)
  \,\biggr]^{p^{ }_0 = k^{ }_{\nu}+q^{ }_{\bar\nu} }_
  {p = |\vec{k}^{ }_{\nu}+\vec{q}^{ }_{\,\bar\nu}|}
 + \rmO(e^2_{ }) 
 \;, \la{Psi_LO}
 \\[2mm]
 \Theta( \vec{q}^{ }_{\nu} \to \vec{k}^{ }_{\nu} )
 & = & 
 + \, \frac{
     G^2_\rmiii{F}
 \, f^{ }_\rmii{B}(p_0^{ })
   }{4\,k^{ }_{\nu}\, q^{ }_{\nu}}
  \Bigl[
    \bigl( 
      2\delta^{ }_{a,e} - 1 + 4 \xW
    \bigr)^2_{ } + 1 
  \Bigr]
 \nn[2mm]
 & \times & 
 \P^2_{ } 
 \, 
  \biggl[\, 
    \rho^\rmii{LO}_\rmii{T} + \rho^\rmii{LO}_\rmii{L} 
  + 
   \biggl( \frac{2 k - p^{ }_0}{p} \biggr)^2_{ }
   \bigl(\, \rho^\rmii{LO}_\rmii{T} - \rho^\rmii{LO}_\rmii{L} \,\bigr)
  \,\biggr]^{p^{ }_0 = k^{ }_{\nu} - q^{ }_{\nu} }_
  {p = |\vec{k}^{ }_{\nu} - \vec{q}^{ }_{\nu}|}
 + \rmO(e^2_{ })
 \;, \la{Theta_LO}
\ea
with $\widetilde\Psi$ given by \eq\nr{balance_1}. 
The leading-order spectral functions can be expressed as 
\ba
  \rho^{\rmii{LO}}_\rmii{T} 
  (p_0,p)
  & = &
  -\; \frac{2 {\cal P}^2 }{p^2} 
  \biggl[ \frac{p_0^2+p^2}2 
  \langle 1 \rangle
  -
  2 \langle r(p_0 - r) \rangle
  \biggr]
  \; , \la{rho_LO_T}
  \\[2mm]
  \rho^{\rmii{LO}}_\rmii{L} 
  (p_0,p)
  & = & 
  -\; \frac{4 {\cal P}^2 }{p^2} 
  \biggl[ - \, \frac{ {\cal P}^2_{ } }{2} 
  \langle 1 \rangle
  +
  2 \langle r(p_0 - r) \rangle
  \biggr] 
  \; , \la{rho_LO_L}
\ea
where
\ba
\langle ... \rangle
& \equiv &
\frac{1}{16 \pi p}
 \biggl\{ 
   \theta({\cal P}^2) \int_{p_-}^{ p_+ } \! {\rm d}r 
 - \theta(-{\cal P}^2)
 \biggl[  \int_{-\infty}^{ p_-} + \int_{p_+}^{\infty} \biggr] 
 \, {\rm d}r
 \biggr\} 
 \bigl[ 1 - \nF^{ }(p_0 - r) - \nF^{ }(r) \bigr]
 (...)
 \; .
 \nn
 \la{spectral_int}
\ea
The integrals over $r$ can be carried out analytically, 
cf.\ \eqs\nr{basis_trafo}--\nr{rho00}. 
Subsequently, 
the integrals over $\vec{q}$ can be carried out numerically, 
with the measures
from \eqs\nr{measure_s} and \nr{measure_t}. The final integral, 
over $\vec{k}$, is radial, 
$
 \int_{\vec{k}^{ }_\nu} = \int_0^\infty \! {\rm d}k^{ }_\nu \, k_\nu^2 
 / (2\pi^2_{ })
$.
The results enter \figs\ref{fig:Q_MB} and \ref{fig:Q}.

%
\subsection{Leading order with Maxwell-Boltzmann statistics}
\la{ss:edot_mb}

It is possible to obtain analytic expressions for the
energy transfer rates, if we make kinematic approximations
in the evaluation of the momentum integrals. Specifically, 
in appendix A.2 of ref.~\cite{mea2}, 
the energy transfer rates 
were computed at LO, 
assuming that 
$f_{\vec{k}^{ }_{\nu}}^{ } = \tilde f_\rmii{F}^{ }(k_\nu^{ })$ and 
$f_{\vec{q}^{ }_{\,\bar\nu}}^{ } = \tilde f_\rmii{F}^{ }(q_{\bar\nu}^{ })$,  
where $\tilde f^{ }_\rmii{F}$ is parametrized by 
a temperature
$T_\nu^{ } \neq T_\gamma^{ }\,$.\footnote{%
  Actually, ref.~\cite{mea2} allows the 
  temperature of each neutrino species 
  to be different. Here we assume that they are the same, i.e.\
  $T_\nu^{ } \equiv T_{\nu_e}^{ } = T_{\nu_\mu}^{ } = T_{\nu_\tau}^{ }\,$,
  but the generalisation is a trivial exercise.
  }
The distribution functions
of both the electron and neutrino ensembles 
were further simplified by
assuming a Maxwell-Boltzmann form, 
$ \tilde f_\rmii{F}^{ }(k_\nu^{ }) \to e^{ - k^{ }_\nu/T^{ }_\nu}_{ }$, 
and neglecting Pauli blocking  
factors, 
$1 - f^{ }_\rmii{F} \to 1$.
We can recover these results in our approach, by 
suitably approximating 
the thermal weights appearing in \eq\nr{spectral_int}.

Specifically, the approximation depends on which arguments of the 
distribution functions are positive. In the end, the 
Maxwell-Boltzmann approximation can be represented as 
the piecewise replacement
\be
 \bigl[ 1 - \nF^{ }(p_0-r) - \nF^{ }(r) \bigr]
 \; = \; 
 \frac{\nF^{ }(p_0-r)\nF^{ }(r)}{\nB^{ }(p_0)}
 \; \stackrel{p^{ }_0 > 0}{\to} \; 
 \left\{
   \begin{array}{lcl}
      e^{r/T^{ }_\gamma} &                &   r < 0 \, \\ 
      1               & \text{\ for\ } &   0 \, < r < p_0^{ } \\
     e^{(p^{ }_0-r)/T^{ }_\gamma}
          &                &   r > p_0^{ } \quad \quad \ \, 
   \end{array} 
   \right.
   . \hspace*{3mm} \la{mb_appro}
\ee

For $s$-channel processes, where 
$0 < r < p^{ }_0$ and $\P^2_{ } > 0$,
\eq\nr{mb_appro} amounts to 
replacing the spectral functions  
by their vacuum limits:
$\rho_\rmii{T} \to - \frac{{\cal P}^2}{12 \pi}$ and 
$\rho_\rmii{L} \to - \frac{{\cal P}^2}{12 \pi}\,$.
Doing so, we obtain
\be
  \Psi(\vec{k}^{ }_{\nu},\vec{q}^{ }_{\,\bar\nu}) 
  \; = \;
    G^2_\rmiii{F} \, 
 \bigl[ ( 2\delta^{ }_{a,e} - 1 + 4 \xW )^2 + 1 \bigr]
  \frac{
    {\cal P}^4 \, e^{-p_0^{ } / T_\gamma^{ } } 
  }{24 \pi \, k \, q} 
  \biggr|^{ p_0^{ } = k+q }_{
    p = |\vec{k}^{ }_\nu + \vec{q}^{ }_{\,\bar\nu} |
  }
 \;.
\ee
We can then evaluate the gain rate from \eq\nr{edot_gain}, 
by first making use of  
\eq\nr{measure_s} to organise the $\vec{q}_{\,\bar\nu}^{ }$ 
integral. 
Thus, neglecting the Pauli blocking factors,
\ba
 \dot{e}^\rmi{gain}_{\nu + \bar\nu}
 & \simeq & 
 \int_{\vec{k}^{ }_\nu,\vec{q}^{ }_{\,\bar\nu}}^{(s)} \hspace*{-1mm}
 \bigl(\, k^{ }_{\nu} +  q^{ }_{\bar\nu} \,\bigr) \;
      \Psi(\vec{k}^{ }_{\nu},\vec{q}^{ }_{\,\bar\nu}) 
 \nn[2mm]
 & = & 
 \frac{
    G^2_\rmiii{F} \, 
 \bigl[ ( 2\delta^{ }_{a,e} - 1 + 4 \xW )^2 + 1 \bigr]
 }{96 \, \pi^5}
 \int_0^\infty \! {\rm d}k
 \int_0^k \! {\rm d}p^{ }_{-}
 \int_k^\infty \! {\rm d}p^{ }_{+} \,  p
 \, p_0^{ } \, 
 {\cal P}^4 \, e^{-p_0^{ } /T_\gamma^{ } }
 \nn[2mm]
 & = & 
 \frac{ G^2_\rmiii{F} }{\pi^5}
 \bigl[ ( 2\delta^{ }_{a,e} - 1 + 4 \xW )^2 + 1 \bigr]
 16 \; T_\gamma^9 
 \; . 
\ea
For annihilation processes, 
we can determine \eq\nr{edot_loss} from detailed balance, 
\ba
 \dot{e}^\rmi{loss}_{\nu + \bar\nu}
 & = & 
 \int_{\vec{k}^{ }_\nu,\vec{q}^{ }_{\,\bar\nu}}^{(s)} \hspace*{-1mm}
 \bigl(\, k^{ }_{\nu} +  q^{ }_{\bar\nu} \,\bigr)
 \bigl[ e^{(k_{\nu}^{ } + q_{\bar\nu}^{ })/T_\gamma}
      \Psi(\vec{k}^{ }_{\nu},\vec{q}^{ }_{\,\bar\nu}) 
  \bigr] 
  e^{- (k_{\nu}^{ } + q_{\bar\nu}^{ })/T_\nu}
 \nn[2mm]
 & = &
 \frac{ G^2_\rmiii{F} }{\pi^5}
 \bigl[ ( 2\delta^{ }_{a,e} - 1 + 4 \xW )^2 + 1 \bigr]
 16 \; T_\nu^9 
 \; .
\ea

For $t$-channel processes, where $\P^2_{ }< 0$, 
the Maxwell-Boltzmann approximated 
spectral functions are more complicated than  
the $s$-channel ones.
For $p_0^{ } >0\,$, the LO replacements are
$\rho_\rmii{T} \to 
  e^{p_{-}^{ } / T_\gamma^{ }} 
 {{\cal P}^2} T_\gamma^{ } 
 (p^2_{ } + 2 p T_\gamma^{ } + 4 T_\gamma^2) / ({4 \pi p^3})$  
and 
$\rho_\rmii{L} \to 
 -\, e^{p_{-}^{ } / T_\gamma^{ }} {{\cal P}^2} T^2_\gamma
 (p + 2  T_\gamma^{ } ) / ({ \pi p^3})$. 
Inserting into \eq\nr{Theta_NLO}, we obtain
\ba
  \Theta(\vec{q}^{ }_{\,\nu}\to\vec{k}^{ }_{\nu}) 
  & = &
    G^2_\rmiii{F} \, 
 \bigl[ ( 2\delta^{ }_{a,e} - 1 + 4 \xW )^2 + 1 \bigr]
  \frac{
    {\cal P}^4 \, e^{ - p^{ }_+ / T_\gamma^{ }} 
  }{16 \pi \, k \, q } 
  \\[2mm]
 & \times &
  \frac{
    T_\gamma \bigl[ 
    p^2(p^2 - 2pT_\gamma - 4T_\gamma^2)
    +
    (2k-p_0)^2 ( p^2 + 6 p T_\gamma + 12 T_\gamma^2 )
    \bigr]
  }{ p^5 } 
  \biggr|^{p_0 = k-q}_{
    p = |\vec{k}^{ }_\nu - \vec{q}^{ }_{\,\nu} |
  }
 \hspace*{-5mm} \;. \hspace*{8mm} \nonumber
\ea
It may be checked that detailed balance is respected as in \eq\nr{balance_2}.
This also guarantees that the domain $p^{ }_0 < 0$ is properly represented, 
even if the approximated $\rho^{ }_\rmii{T,L}$ do not explicitly
manifest the asymmetry
$\rho_i^{ }(p_0^{ }) = - \rho_i^{ } (-p_0^{ })\,$. 
The reason for the missing symmetry is that
Maxwell-Boltzmann statistics is not accurate when 
$|p^{ }_0| \ll \pi T$. 

Using \eq\nr{measure_t},  
we can now evaluate \eq\nr{edot_scat}. Even though the intermediate
steps are quite complicated, owing to the appearance of 
two different temperatures, the result simplifies dramatically once
the integral over $\vec{q}^{ }_{\,\nu}$ is carried out. After the remaining
integral over $\vec{k}^{ }_\nu$, we obtain
\ba
 \dot{e}^\rmi{scat}_{\bar \nu \to \bar \nu} 
 \ \  = \ \ 
 \dot{e}^\rmi{scat}_{\nu \to \nu}
 & = &
 \int_{\vec{k}^{ }_\nu,\vec{q}^{ }_{\,\nu}}^{(t)} \hspace*{-1mm}
 \bigl(\, k^{ }_{\nu} -  q^{ }_{\nu} \,\bigr) \;
      {\Theta}(\vec{q}^{ }_{\,\nu} \to \vec{k}^{ }_{\nu})  \;
      e^{-q/T_\nu}
  \nn[2mm]
 & = &
 \frac{ G^2_\rmiii{F} }{\pi^5}
 \bigl[ ( 2\delta^{ }_{a,e} - 1 + 4 \xW )^2 + 1 \bigr]
 14 \; 
 T_\gamma^4 \; 
 T_\nu^4 \; 
 \bigl( T_\gamma - T_\nu \bigr)
 \; .
\ea
Combining all the individual energy transfer rates, we find 
\be
  Q^\rmii{MB}_{ } \; = \; 
 \frac{ G^2_\rmiii{F} }{\pi^5}
 \bigl[ ( 2\delta^{ }_{a,e} - 1 + 4 \xW )^2 + 1 \bigr]
 \Bigl\{
   16 \bigl( T_\gamma^9 - T_\nu^9 \bigr) 
   \; + \; 
 28 \; 
 T_\gamma^4 \; 
 T_\nu^4 \; 
 \bigl( T_\gamma - T_\nu \bigr)
 \Bigr\} 
 \; . \la{Q_res}
\ee

%
\begin{figure}[t]

\centerline{
     \epsfysize=7.0cm\epsfbox{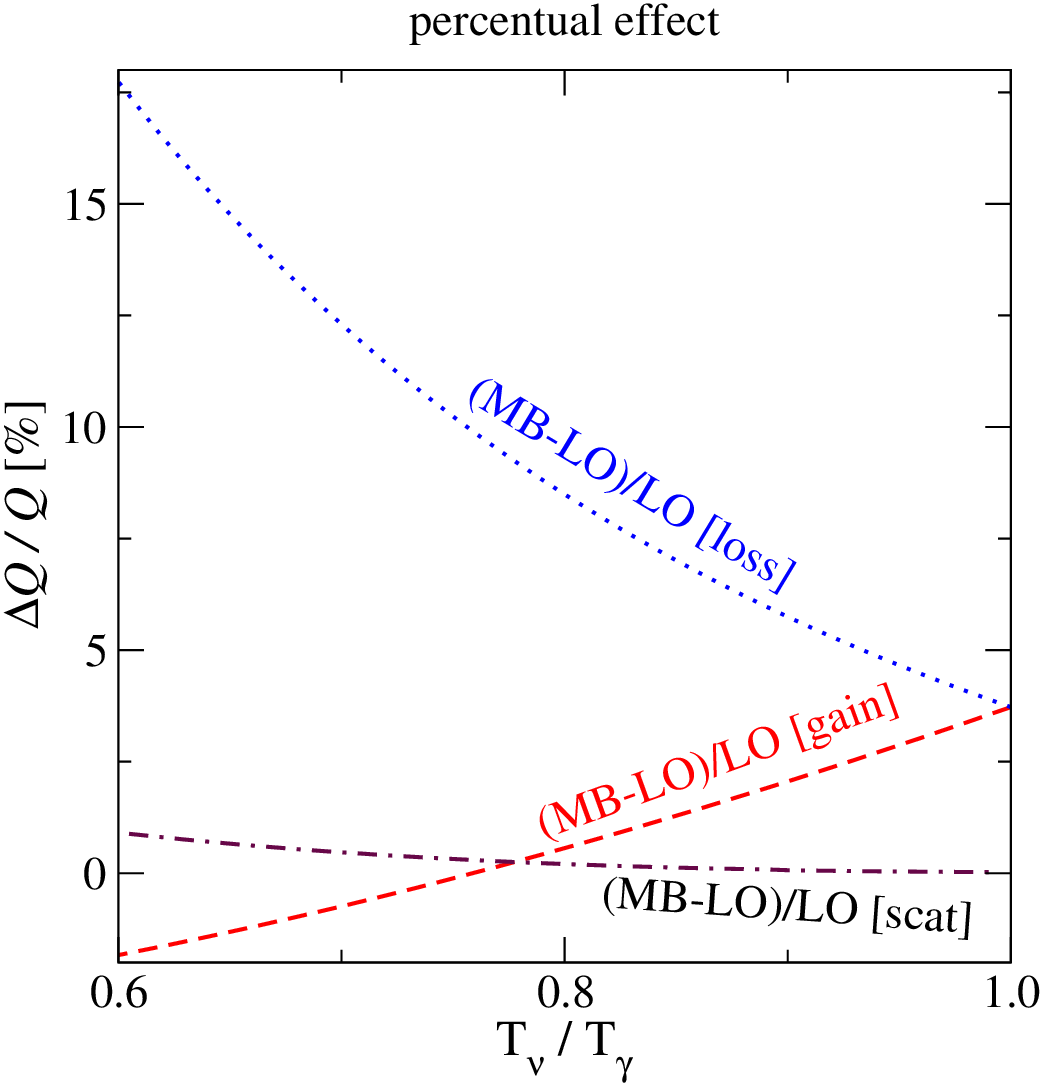}
}

\vspace*{1mm}

\caption[a]{\small 
 Percentual accuracy of 
 the improved Maxwell-Boltzmann approximation from
 \eq\nr{Q_mb} (MB),
 compared with  the full LO evaluation based on 
 \eqs\nr{Psi_LO}--\nr{spectral_int} (LO), 
 on the 
 energy transfer rates from \eqs\nr{edot_gain}--\nr{edot_scat}. 
 The results are similar to those found in ref.~\cite{eks}.
} 
\la{fig:Q_MB}
\end{figure}
%

Equation~\nr{Q_res}
agrees with eq.~(A.15) of ref.~\cite{mea2}, 
where the same result was obtained by taking 
moments of the Boltzmann equation. 
In the actual neutrino decoupling computation
of ref.~\cite{cemp}, 
the coefficients were corrected in order to account
for Fermi-Dirac statistics, as 
\be
  Q^\rmii{MB}_{ } \; \to \; 
 \frac{ G^2_\rmiii{F} }{\pi^5}
 \bigl[ ( 2\delta^{ }_{a,e} - 1 + 4 \xW )^2 + 1 \bigr]
 \Bigl\{
 16 \, f^\rmii{FD}_{a}
 \, 
 \bigl( \underbrace{ T_\gamma^9 }_{\rm gain} -
        \underbrace{ T_\nu^9 }_{\rm loss}  \bigr) 
   \; + \; 
 28 \, f^\rmii{FD}_{s}
 \,
 \underbrace{ 
 T_\gamma^4 \; 
 T_\nu^4 \; 
 \bigl( T_\gamma - T_\nu \bigr) }_{\rm scat}
 \Bigr\} 
 \; , \la{Q_mb}
\ee
where the numerically determined factors 
$f_a^\rmii{FD} = 0.884$ (``annihilations'') 
and $f_s^\rmii{FD} = 0.829$ (``scatterings'')
guarantee that 
$Q^\rmii{MB}_\rmi{gain} - Q^\rmii{MB}_\rmi{loss}$ 
and $Q^\rmii{MB}_\rmi{scat}$ 
approach zero as $T^{ }_\nu\to T^{ }_\gamma$ with the same slopes as 
$Q^\rmii{LO}_\rmi{gain} - Q^\rmii{LO}_\rmi{loss}$ 
and $Q^\rmii{LO}_\rmi{scat}$, respectively.
In \fig\ref{fig:Q_MB}, we compare \eq\nr{Q_mb}
with the exact LO rates originating from 
\eqs\nr{Psi_LO}--\nr{spectral_int}.

%
\subsection{Full next-to-leading order}
\la{ss:edot_nlo}

The NLO expressions for the double-differential rates appearing 
in \eqs\nr{edot_gain}--\nr{edot_scat} are given in 
\eqs\nr{balance_1}, \nr{Psi_NLO} and \nr{Theta_NLO}. The key 
ingredient entering the double-different rates is the 
function $\mathcal{F}$, given in \eq\nr{calF}. 
It captures two aspects of the physics being considered: 
on one hand, coupling constants
and flavour factors, the latter originating from the fact that
electron-flavoured neutrinos experience different QED corrections
than $\mu$ and $\tau$-flavoured ones. On the other hand, there
is the dynamical information, originating from matrix elements
squared associated with the reactions taking place, 
and phase-space integrals over the thermal electrons
and positrons. This dynamical information is 
captured by the spectral functions 
$\rho^{ }_\rmii{T,L}$, vertex corrections $\chi^{ }_\rmii{T,L}$, 
and resummation factors $\mathcal{R}^*_\rmii{T,L}$.

In order to factorize the two different ingredients 
affecting the energy transfer rates, it is helpful to 
pull out the flavour factors and the QED coupling constant
in front of the dynamical information. To this aim, we can 
rewrite \eq\nr{calF} as 
\be
 \mathcal{F}(k;p^{ }_0,p) 
 \; \equiv \; 
 T^4_{\gamma} \, 
 \biggl\{\, 
   \widehat{\mathcal{F}}^{ }_\rmii{T}(p^{ }_0,p) 
  + 
   \widehat{\mathcal{F}}^{ }_\rmii{L}(p^{ }_0,p) 
  + 
   \biggl( \frac{2k - p^{ }_0}{p} \biggr)^2_{ } 
   \bigl[
   \widehat{\mathcal{F}}^{ }_\rmii{T}(p^{ }_0,p) 
  -
   \widehat{\mathcal{F}}^{ }_\rmii{L}(p^{ }_0,p)  
   \bigr]
 \,\biggr\}
 \;. \la{calF_rep}
\ee
Here the dimensionless coefficients read, 
for $i = \rmi{T,L}$,
\ba
 \widehat{\mathcal{F}}^{ }_i 
 & = & 
 \frac{\P^2_{ }}{T^4_{\gamma} } \, \Bigl\{ \, 
   \bigl[
    \bigl( 
      2\delta^{ }_{a,e} - 1 + 4 \xW
    \bigr)^2_{ } + 1 
  \bigr]
  \rho^\rmii{NLO}_{i}
 + 
 e^2_{ } C^{ }_a 
     \bigl( 
      2\delta^{ }_{a,e} - 1 + 4 \xW
    \bigr)
  \rho^\rmii{LO}_{i}
\nn[2mm]
 &  & +  \,
 2 e^2_{ }
     \bigl( 
      2\delta^{ }_{a,e} - 1 + 4 \xW
    \bigr)^2_{ }
 \rho^\rmii{LO}_i \chi^\rmii{LO}_i \mathcal{R}^*_i 
 \Bigr\}
 \;. 
\ea
The NLO spectral functions are written as 
\be
  \rho^\rmii{NLO}_{i}
 \; = \; 
   \rho^\rmii{LO}_{i}
 + 
  e^2_{ } 
 \delta   \rho^\rmii{NLO}_{i}
 \;, 
\ee
where $ \delta \rho^\rmii{NLO}_{i} $ were determined in 
refs.~\cite{twoloop,twoloop_code}. 
The coefficient $C^{ }_a$ combines a counterterm and a matching
coefficient, and can according to \eq(2.17) of  
ref.~\cite{rate} be expressed as 
\be
 \frac{ e^2_{ }C^{ }_a }{8} \biggr|^{ }_\rmi{bare} 
 \; = \;
 ( 2\delta^{ }_{a,e} - 1 + 4 \xW )
 \frac{e^2}{3} \frac{\mu^{-2\epsilon}}{(4\pi)^2} 
 \biggl( \frac{1}{\epsilon} + \ln\frac{\bmu^2}{m_e^2} \biggr)
 \; + \;
 (-0.01 \, ... \, 0.01)
 \;, \la{C_a}
\ee
where the uncertainty originates from low-energy hadronic input~\cite{eft2}. 
The functions $\chi^\rmii{LO}_i$ represent the real parts of closed
electron-positron-loops, and can be written as 
\be
 \chi^\rmii{LO}_i 
 \; = \; 
 \chi^\rmii{LO}_i \bigr|^{ }_\rmi{vac} 
 + 
 \chi^\rmii{LO}_i \bigr|^{ }_{\T_{\!\gamma}}
 \;, \la{chi_splitup}
\ee
where the vacuum parts are 
\be
 \chi^\rmii{LO}_i \;\bigr|^{ }_\rmi{vac}
 \; = \; 
 - \frac{4\P^2_{ }}{3}
   \frac{\mu^{-2\epsilon}_{ }}{(4\pi)^2}
  \biggl(
    \frac{1}{\epsilon} + \ln\frac{\bmu^2_{ }}{|\P^2_{ }|} + \fr53
  \biggr)
 \;. \la{chi_vac}
\ee
The divergence from here cancels against that in \eq\nr{C_a}. 
The thermal parts are given in \eqs\nr{chiV} and \nr{chi00}.
Finally, 
$\mathcal{R}^*_i$ represent the resummed photon propagator
(cf.\ \eq\nr{resum}), normalized
so that in the domain requiring no resummation, 
\be
 \mathcal{R}^*_i
 \overset{|\P^2| \gg e^2 T^2}{\longrightarrow}
 \mathcal{R}^{ }_i
 \; \equiv \; \frac{1}{\P^2_{ }}
 \;. 
\ee
Resummation is only needed in $t$-channel processes enhanced by thermal
effects, and therefore we may write 
\be
 \chi^\rmii{LO}_i \mathcal{R}^*_i
 \; \approx \; 
 \chi^\rmii{LO}_i \bigr|^{ }_\rmi{vac} \mathcal{R}^{ }_i 
 + 
 \chi^\rmii{LO}_i \bigr|^{ }_{\T_{\!\gamma}}
 \, \bigl[ 
   \theta(\P^2_{ }) \mathcal{R}^{ }_i 
 +  
   \theta(-\P^2_{ }) \mathcal{R}^{*}_i
 \bigr]
 \;.  
\ee
Putting all of these ingredients together, we find 
\ba
 \widehat{\mathcal{F}}^{ }_i
 & \equiv & 
 \bigl[ \,
   \bigl( 2\delta^{ }_{a,e} - 1 + 4 \xW \bigr)^2_{ }
 + 1 
 \, \bigr] \,
 \biggl\{\, 
   \underbrace{ 
   \biggl( \frac{\P^2_{ } \rho^\rmii{LO}_i}{T^4_{\gamma}} \biggr)
   }_{ 
   \,\equiv\, \A^{ }_i }
   + 
   e^2_{ }\,  
   \underbrace{ 
   \biggl( \frac{\P^2_{ } \delta \rho^\rmii{NLO}_i}{T^4_{\gamma}} \biggr)
   }_{ 
   \,\equiv\, \B^{ }_i }
 \,\biggr\} 
 \nn[2mm] 
 & &  +  \,
 e^2_{ } \, \bigl( 2\delta^{ }_{a,e} - 1 + 4 \xW \bigr)^2_{ } \, 
 \biggl\{ 
   \frac{1}{6\pi^2}    
   \biggl[
   \underbrace{ 
   \biggl( \frac{\P^2_{ } \rho^\rmii{LO}_i 
           \ln| \P^2_{ }/ T^2_{\gamma} | }{T^4_{\gamma}} \biggr)
   }_{ 
   \,\equiv\, \C^{ }_i }
        - \biggl( 2\ln\frac{m^{ }_e}{T^{ }_{\gamma}} + \frac{5}{3} \biggr)
   \, 
   \underbrace{ 
   \biggl( \frac{\P^2_{ } \rho^\rmii{LO}_i}{T^4_{\gamma}} \biggr)
   }_{ 
   \, = \, \A^{ }_i }
   \biggr]
 \biggr\} 
 \nn[2mm] 
 & &  +  \,
 e^2_{ } \, \bigl( 2\delta^{ }_{a,e} - 1 + 4 \xW \bigr)^2_{ } \, 
 \biggl\{ 
 \underbrace{
 \biggl( 
   \frac{2 \rho^\rmii{LO}_i \chi^\rmii{LO}_i |^{ }_{\T_{\!\gamma}} }
 {T^4_{\gamma}}
 \biggr)
 \bigl[\,
   \theta(\P^2_{ }) + 
   \theta(-\P^2_{ }) \P^2_{ }\mathcal{R}^*_i
 \,\bigr]
 }_{ 
  \,\equiv\,\D^{ }_i
 }
 \biggr\} 
 \nn[2mm] 
 & & + \, 
 e^2_{ } \, \bigl( 2\delta^{ }_{a,e} - 1 + 4 \xW \bigr)^{ }_{ } \, 
 \biggl\{ 
     \underbrace{8\, (-0.01 \, ... \, 0.01)  
     }_{\rm from~uncertainty~of~{\it C^{ }_a}} 
   \underbrace{ 
   \biggl( \frac{\P^2_{ } \rho^\rmii{LO}_i}{T^4_{\gamma}} \biggr)
   }_{ 
   \, = \, \A^{ }_i }
 \biggr\} 
 + \rmO(e^3_{ })
 \;. \la{ABCD}
\ea

To summarize, the complete function $\mathcal{F}$ from 
\eqs\nr{calF} and \nr{calF_rep}, including its uncertainty
and $k$-dependence, 
can be reconstructed, once we know the coefficients
$\A^{ }_i$,
$\B^{ }_i$,
$\C^{ }_i$ and
$\D^{ }_i$, with $i = \rmi{T,L}$.
We note that the coupling $e^2_{ } = 4\pi \alpha^{ }_\rmi{em}$, 
with $\alpha^{ }_\rmi{em} = 1/137$, has been factored out 
from these coefficients, however it still appears 
inside the resummation factors $\mathcal{R}^*_i$
(cf.\ \eq\nr{resum}), 
which influence $\D^{ }_i$ in the spacelike 
region.

Once we insert \eq\nr{ABCD} into \eq\nr{calF_rep}, 
and then \eq\nr{calF_rep} into \eqs\nr{Psi_NLO}, \nr{Theta_NLO},
and \nr{balance_1}, and then these 
into \eqs\nr{edot_gain}--\nr{edot_scat}, we obtain NLO 
results for the various energy transfer rates. The latter
will be denoted by $Q^\rmii{NLO}_{ }$. Following the
structures of \eqs\nr{calF_rep} and \nr{ABCD}, it is natural
to express the results as 
\ba
 \frac{Q^\rmii{NLO}_{ }}{ G^2_\rmiii{F} T_\gamma^9 } 
  & = &
 \bigl[ \,
   \bigl( 2\delta^{ }_{a,e} - 1 + 4 \xW \bigr)^2_{ }
 + 1 
 \, \bigr] \,
 \biggl\{\, 
   \widehat Q_{_\A} 
   + 
   e^2_{ }\,  
   \widehat Q_{_\B}  
 \,\biggr\} 
 \nn[2mm] 
 & + &  
 e^2_{ } \, \bigl( 2\delta^{ }_{a,e} - 1 + 4 \xW \bigr)^2_{ } \, 
 \biggl\{ 
   \frac{1}{6\pi^2}    
   \biggl[
     \widehat Q_{_\C}
        - \biggl( 2\ln\frac{m^{ }_e}{T^{ }_{\gamma}} + \frac{5}{3} \biggr)
   \, 
     \widehat Q_{_\A}
   \biggr]
 \, + \, 
 \widehat Q_{_\D}
 \biggr\} 
 \nn[2mm] 
 & + &  
 e^2_{ } \, \bigl( 2\delta^{ }_{a,e} - 1 + 4 \xW \bigr) \, 
 \biggl\{ 
     8\, (-0.01 \, ... \, 0.01)  
     \widehat Q_{_\A} 
 \biggr\}
 \;, \la{overall_Q}
\ea
where 
  $\widehat Q_{_\A}$,
  $\widehat Q_{_\B}$,
  $\widehat Q_{_\C}$, and 
  $\widehat Q_{_\D}$ 
are dimensionless functions of $T_\nu^{ }/T_\gamma^{ }\,$, 
which originate from the coefficients 
$\A^{ }_i$, $\B^{ }_i$, $\C^{ }_i$, and $\D^{ }_i$, respectively. 

Apart from the dependence of
  $\widehat Q_{_\A}$,
  $\widehat Q_{_\B}$,
  $\widehat Q_{_\C}$,  
  $\widehat Q_{_\D}$ 
on $T_\nu^{ }/T_\gamma^{ }\,$,
the only other temperature dependence in \eq\nr{overall_Q}
is from $\ln( m_e^{ }/T_\gamma^{ }\,)$. 
We remark that the representation
in \eq\nr{overall_Q}
is reliable in the domain $m^{ }_{e} \le T^{ }_\gamma \ll m^{ }_{\mu}$.
In contrast, if $T^{ }_\gamma < m^{ }_e$, the spectral functions, 
vertex corrections, and resummation factors, obtain an additional
non-trivial dependence on $m^{ }_e/T^{ }_\gamma$, 
which is not included in our NLO results. The full dependence 
on $m^{ }_e/T^{ }_\gamma$ is worked out at leading order in appendix~\ref{app:B}.

The last line of \eq\nr{overall_Q} includes the uncertainty
from \eq\nr{ABCD}, originating from poorly determined low-energy 
hadronic input to the Fermi effective theory. Normalizing 
to leading-order results, the uncertainty can be expressed as 
\be
 \frac{\delta Q^\rmii{NLO}_{ }}{Q^\rmiii{LO}_{ }}
 \; = \; 
 \frac{
  \bigl|
 e^2_{ } \, \bigl( 2\delta^{ }_{a,e} - 1 + 4 \xW \bigr)^{ }_{ } \, 
 \bigl\{ 
     {8\, (\,0.01\,)}
 \bigr\} 
  \bigr|
 }{
 \bigl[ \,
   \bigl( 2\delta^{ }_{a,e} - 1 + 4 \xW \bigr)^2_{ }
 + 1 
 \, \bigr] 
 }
 \; = \; 
 \left\{
   \begin{array}{l}
      0.2976 \ \% \\
      0.0334 \ \% \\
   \end{array} 
   \right.
   \text{\ for\ }
 \left.
   \begin{array}{l}
      a = e  \\
      a \neq e
   \end{array} 
   \right. 
 \;. 
 \la{uncertainty}
\ee

For a practical determination of 
  $\widehat Q_{_\A}$,
  $\widehat Q_{_\B}$,
  $\widehat Q_{_\C}$, and 
  $\widehat Q_{_\D}$, 
we precompute 
$\A^{ }_i$, $\B^{ }_i$, $\C^{ }_i$, and $\D^{ }_i$
on a grid, and then interpolate. 
The grid and the interpolation routine are described
in \se\ref{se:tabulation}. Subsequently, 
we carry out the remaining 3-dimensional integral numerically, 
first over the momentum transfer according to \eq\nr{measure_s} or 
\nr{measure_t}, and subsequently over $k^{ }_\nu$.

%
\begin{figure}[t]

\centerline{
     \epsfysize=8.3cm\epsfbox{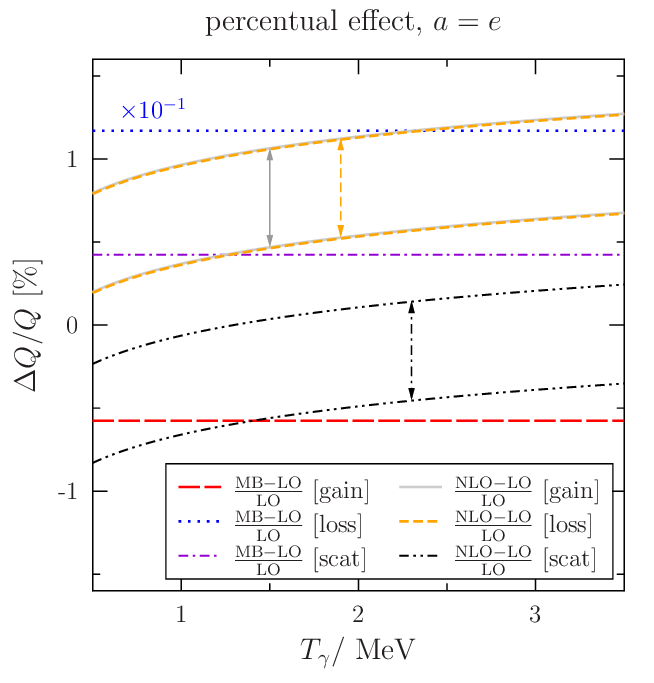}
    ~\epsfysize=8.3cm\epsfbox{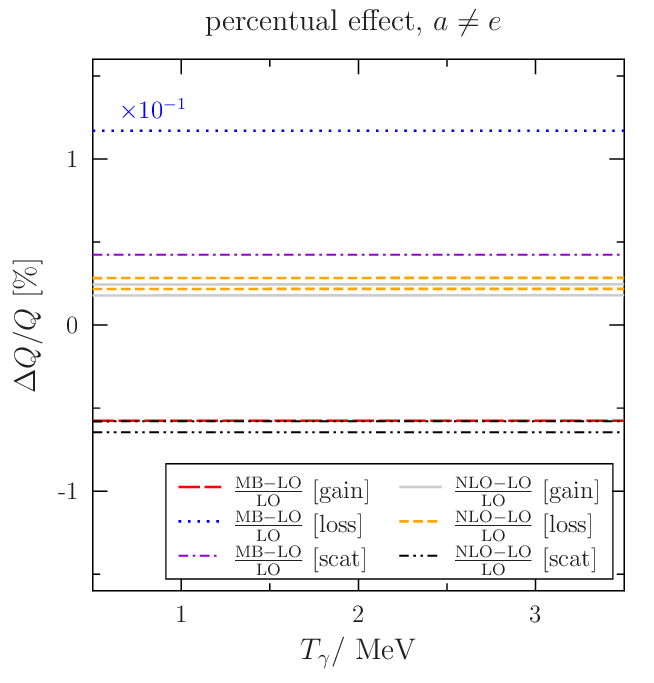}
}

\vspace*{-2mm}

\caption[a]{\small 
 Percentual influence of various
 effects on the 
 energy transfer rates from \eqs\nr{edot_gain}--\nr{edot_scat}. 
 ``MB'' refers to the improved Maxwell-Boltzmann approximation from
 \eq\nr{Q_mb}, ``LO'' to the full LO evaluation based on 
 \eqs\nr{Psi_LO}--\nr{spectral_int}, and ``NLO'' to the 
 full NLO result originating from 
 \eqs\nr{Psi_NLO} and \nr{Theta_NLO}.
 The arrows show the uncertainly from low-energy hadronic input, 
 as specified in \eq\nr{uncertainty}.  
 For illustration we have set 
 $
 \xW |^{ }_{\bmu = m^{ }_e} \simeq 0.2386
 $
 and 
 $
  T^{ }_\nu = (4/11)^{1/3}_{ }T^{ }_\gamma 
  \approx 
  0.714\,T^{ }_\gamma
 $.
 At this temperature, the MB approximation works fairly well
 for the gain and scattering terms (cf.\ \fig\ref{fig:Q_MB}), but 
 poorly for the loss term, whence we have multiplied it with 
 $10^{-1}_{ }$. 
 The NLO gain and loss terms are practically independent
 of $T^{ }_\nu/T^{ }_\gamma$, whereas the scattering
 terms show a mild variation. 
} 
\la{fig:Q}
\end{figure}
%

Our results for $Q^\rmii{NLO}_{ }$, 
compared with the leading-order values from 
\se\ref{ss:edot_lo}, are shown in \fig\ref{fig:Q}. 
We plot the results
as a function of $T^{ }_\gamma$, 
assuming for illustration
$
 T^{ }_\nu = (4/11)^{1/3}_{ }T^{ }_\gamma \approx 0.714\, T^{ }_\gamma
$.
The decoupling dynamics takes place between this minimal value, 
and the maximal value $T^{ }_\nu = T^{ }_\gamma$; we have checked
that our results are stable within 
this range of $T^{ }_\nu/T^{ }_\gamma$, in fact also if 
$T^{ }_\nu \ll T^{ }_\gamma$. 
The NLO corrections are generally $\sim \pm 1\%$. 
For the gain and scattering terms, they are similar 
in magnitude to the error of 
the MB approximation at $T^{ }_\nu  \approx 0.714\, T^{ }_\gamma$,
however the MB error increases when
going towards  $T^{ }_\nu = T^{ }_\gamma$
(cf.\ \fig\ref{fig:Q_MB}). 
For the loss term, the NLO correction is an order of 
magnitude smaller than the error from 
the MB approximation. What all of this
implies for $N_\rmi{eff}$ will be
elaborated upon in \se\ref{se:conclu}.

%
\section{Tabulation of full double-differential rates in the massless limit}
\la{se:tabulation}

A full NLO computation of $N^{ }_\rmi{eff}$ requires the solution 
of quantum kinetic equations as a function of the neutrino spatial 
momentum, $k$. Indeed it is believed that neutrinos deviated from
equilibrium in a momentum-dependent manner
(for illustrations see,\ e.g.,\ fig.~3 of ref.~\cite{Neffm2} 
or fig.~4 of ref.~\cite{Neffm1}). Momentum-dependent equations 
are parametrized by momentum-dependent rates, rather than the 
averaged ones in \eqs\nr{edot_gain}--\nr{edot_scat}. In this 
section, we tabulate the full momentum-dependent 
double-differential pair-production, 
annihilation, and scattering rates at NLO. We stress that these functions
have been determined without any kinematic approximations, 
after a careful choice of integration variables had permitted to 
make the remaining numerical effort manageable enough to be 
 recorded on a dense grid (see below). 

The momentum-dependent information is contained in the function 
$\mathcal{F}$, defined in \eq\nr{calF}. 
It determines the gain terms according to \eq\nr{Psi_NLO}, 
the loss terms according to \eq\nr{balance_1}, and the 
scattering terms according to \eq\nr{Theta_NLO}.
In the $s$-channel contribution, relevant for the gain and loss terms, 
the four-momentum transfer, $\P$, is
in the timelike domain, whereas in the $t$-channel contribution, 
relevant for the scattering terms, 
it is in the spacelike domain.

%
\begin{figure}[t]

\centerline{
     \epsfysize=7.0cm\epsfbox{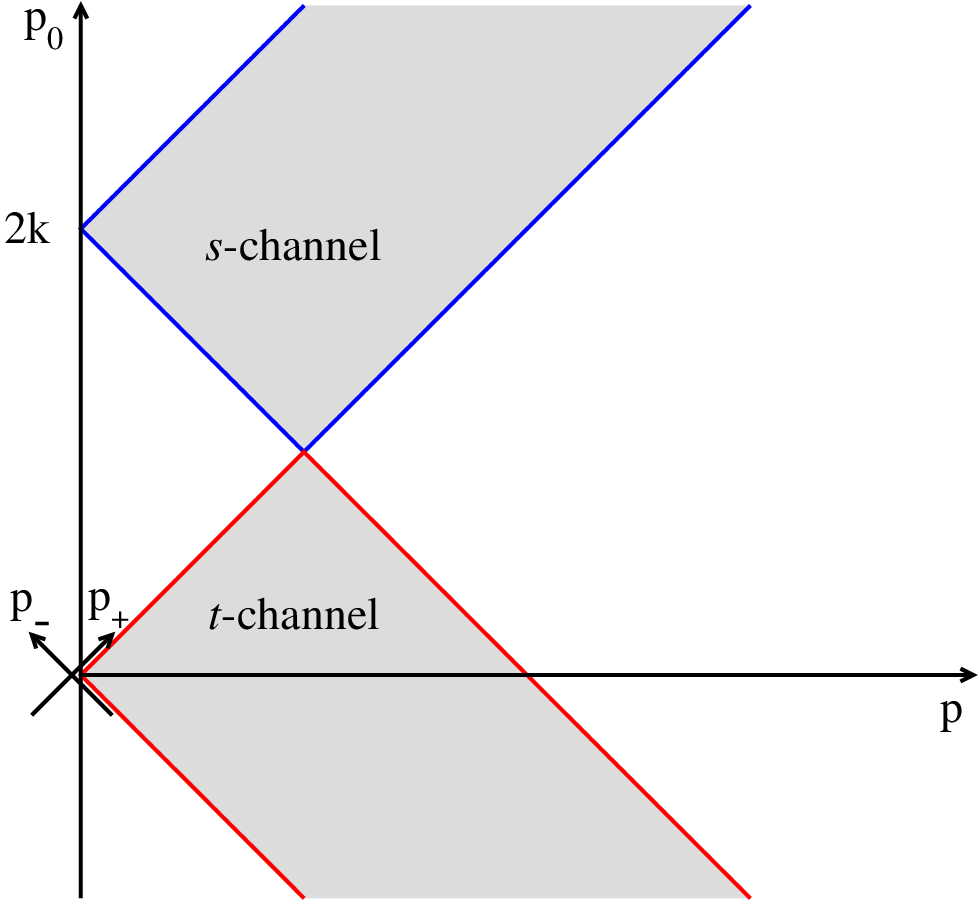}
}

\vspace*{1mm}

\caption[a]{\small 
 The domains of four-momentum transfer that are relevant 
 for a given neutrino spatial momentum $k$, as found in 
 \eqs\nr{measure_s} and \nr{measure_t}. The spectral functions
 are antisymmetric in $p^{ }_0\to - p^{ }_0$, so the full information
 is already contained in the first quadrant $p^{ }_0 > 0$.
} 
\la{fig:kinematics}
\end{figure}
%

It is important to note that the 
neutrino spatial momentum, $k$, 
is an auxiliary variable from the point of view of the four-momentum
transfer. It appears as a weighting factor in \eq\nr{calF}, and 
in addition it determines the integration ranges in 
\eqs\nr{measure_s} and \nr{measure_t}. However, it does not appear
inside the functions $\rho$, $\chi$, $\mathcal{R}^*_{  }$ that
determine the value of \eq\nr{calF}. The kinematic domains of 
$p^{ }_0$ and $p$ that are realized for a given $k$ are 
illustrated in \fig\ref{fig:kinematics}.

We therefore precompute the coefficients 
$\A^{ }_i$, 
$\B^{ }_i$, 
$\C^{ }_i$ and 
$\D^{ }_i$, 
in the first quadrant of the $(p,p^{ }_0)$--plane.
Exemplary value are illustrated in table~\ref{table:coeffs},  
however for the actual interpolation a much denser grid
is constructed.
{}From the first quadrant, 
we can recover coefficients 
in all the domains of \fig\ref{fig:kinematics}, 
due to 
general properties of the photon self-energy, 
namely 
$\rho_i^{ }(p_0^{ }) = - \rho_i^{ } (-p_0^{ })$ and 
$\chi_i^{ }(p_0^{ }) = \chi_i^{ } (-p_0^{ })\,$. 

In practice, 
we employ a wedge-shaped grid, illustrated 
in \fig\ref{fig:grid} (blue circles), 
which is uniform in the rotated coordinates $(p^{ }_+,p^{ }_-)$, 
with a ``lattice spacing'' of $\delta\,$.
If we consider the grid to have $N$ points falling 
on the $p^{ }_+$ axis,
the maximal value of $p^{ }_+$ on the grid is 
$p^{ }_{\rm max} = N \delta $
and the minimal value of $p^{ }_+$ is  $p^{ }_{\rm min} = \delta\,$.
The grid comprises a total of $N^2_{ }$ points.

%
\begin{figure}[t]

\centerline{
     \epsfysize=7.0cm\epsfbox{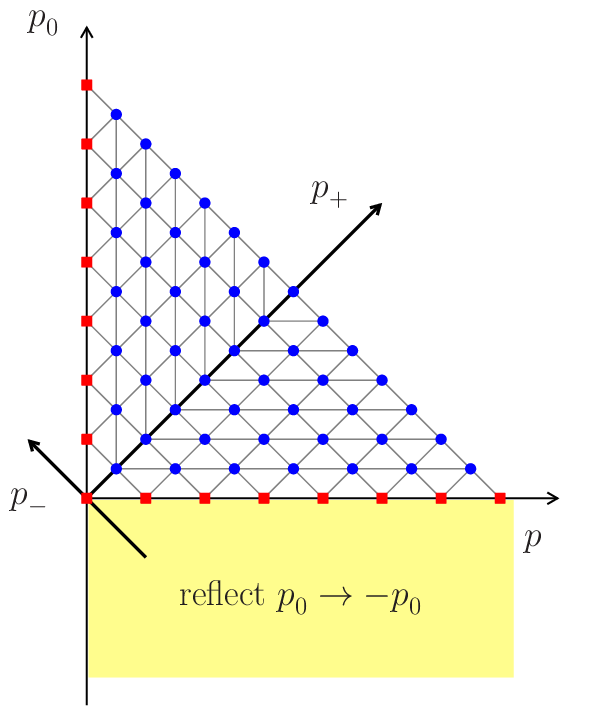}
}

\vspace*{1mm}

\caption[a]{\small 
 Grid for tabulation (circles and squares) and interpolation schemes.
 At the blue circles, 
 we record the coefficients in \eq\nr{ABCD}.
 The red squares ``pad'' the boundary, and are useful because 
 the factor in
 \eq\nr{interpolant} results in the integrand being zero on these points
  (even if the coefficients are non-trivial).
 The shaded region 
 may be obtained by reflecting $p^{ }_0 \to -p^{ }_0$ and 
 using antisymmetry of the spectral functions. The coefficients
 turn out to vanish also at $p^{ }_- = 0\,$, where $\P^2_{ }= 0$, 
 however the approach to this line is not smooth: without 
 further resummations, which have not been implemented in this
 work, the coefficient of $\P^2_{ }$ in $\B^{ }_i$ is logarithmically 
 divergent and discontinuous as $\P^2_{ } \to 0$. 
} 
\la{fig:grid}
\end{figure}
%

The idea is then to interpolate, from the tabulated values, the 
$p_0$ and $p$ dependence of the integrand for the quantity of 
interest 
(for example, 
the energy density transfer rates 
defined in \se\ref{se:energy}).
To accomplish this, we augment the (blue) recorded circles 
with the (red) boundary squares shown in \fig\ref{fig:grid}.
There are $(N+1)^2_{ }$ of these points altogether, from 
which we can triangulate the grid, 
and interpolate the value at a given $(p_+^{ },p_-^{ })$  
as a linear combination of the three vertices of the triangle 
in which the point falls. 
Rather than interpolating the coefficients themselves, 
which may have peculiarities at the 
boundaries (for $p = 0$ or $p_0^{ }=0$), 
we instead interpolate
\be
 \nB^{ }(p^{ }_0) \, 
 p^{ }_0 \, 
 p \, 
 \mathcal{F}(k;p^{ }_0,p) 
 \;, 
 \la{interpolant}
\ee
which appears as a common integration kernel when 
computing the contributions to $Q\,$
(cf.\ \eqs\nr{measure_s}, \nr{measure_t}, and
\nr{edot_gain}--\nr{edot_scat}). 
The quantity in \eq\nr{interpolant} is also 
convenient for being zero at the boundaries.\footnote{%
  For $p^{ }_0=0\,$, the spectral functions are zero due 
  to antisymmetry in their energy argument, 
  which results in $\mathcal{F}$ also being zero there.
  In the limit $p\to 0\,$, we observe that 
  $(\rho^{ }_\rmii{T}-\rho^{ }_\rmii{L}) \sim p^2_{ }$
  compensates the factor $({2k - p^{ }_0})^2_{ }/{p}^2_{ }$
  appearing in \eq\nr{calF_rep}, 
  so that \eq\nr{interpolant} is similarly zero.
  }

To finally carry out the integration in 
\eqs\nr{edot_gain}, 
\nr{edot_loss} and 
\nr{edot_scat}, 
we should restrict the infinite integration ranges,
\ba
 \int_0^\infty {\rm d} k 
 \int_0^{k} {\rm d} p^{ }_-
 \int_{k}^\infty {\rm d} p^{ }_+
  & \rightarrow &
 \int_0^{p_{\rm max}} {\rm d} k 
 \int_0^{k} {\rm d} p^{ }_-
 \int_{k}^{p_{\rm max}} {\rm d} p^{ }_+ 
 \  ,
 \\
 \int_0^\infty {\rm d} k 
 \int_{-\infty}^0 {\rm d} p^{ }_-
 \int_0^{k} {\rm d} p^{ }_+
  & \rightarrow &
 \int_0^{p_{\rm max}} {\rm d} k 
 \int_{-p_{\rm max}}^0 {\rm d} p^{ }_-
 \int_0^{k} {\rm d} p^{ }_+
 \  .
\ea
This numerical integration is still somewhat expensive, but 
using a ``look-up table'' for the integrand is desirable at NLO because
 $\rho_i^\rmii{NLO}$ itself originates from 
 a costly two-dimensional integral. 
The integrand seems to be well behaved 
(including at the boundaries) in all cases. 
For the numerical NLO results shown in \fig\ref{fig:Q}, 
we have used $N=500$ and $p_{\rm max} = 25 \, T_\gamma\,$. 
This table reproduces 
the full $Q^\rmii{LO}_{ }$ with an accuracy of $< 0.05$\%. 
For convenience, 
we make this data as well as an interpolation routine 
publicly available~\cite{interpolation_code}. 
We have checked that the table can also be used to reconstruct the 
neutrino interaction rate\hspace*{0.3mm}\footnote{%
 For the neutrino interaction rate,
 it is preferable to interpolate the function 
 \be
 \big[ 1 - \nF^{ }(k-p^{ }_0) + \nB^{ }(p^{ }_0) \big]
 \, p \, 
 \mathcal{F}(k;p^{ }_0,p) 
 \;, 
 \ee
 instead of \eq\nr{interpolant}.
 Integrating over $p^{ }_+$ and $p^{ }_-$ according to 
 \eqs\nr{measure_s} and \nr{measure_t} then allows 
 for the coefficients $A$, $B$, $C$, $D$ 
 of ref.~\cite{rate} 
 to be reconstructed.
}
computed in ref.~\cite{rate}. 

%
\begin{table}[t]

{\fontsize{8pt}{10pt}\selectfont
$$
\begin{array}{rrrrrrrrrr} 
  p / T_\gamma^{ } & p_0^{ } / T_\gamma^{ }
  & \mathcal{A}^{ }_\rmii{T} 
  & \mathcal{A}^{ }_\rmii{L} 
  & \mathcal{B}^{ }_\rmii{T} 
  & \mathcal{B}^{ }_\rmii{L} 
  & \mathcal{C}^{ }_\rmii{T} 
  & \mathcal{C}^{ }_\rmii{L} 
  & \mathcal{D}^{ }_\rmii{T} 
  & \mathcal{D}^{ }_\rmii{L} 
 \\[1mm]
  \hline \\[-3mm]
  0.10 & 0.05 &
  +0.737m3  & 
  -0.147m2  & 
  -0.699m4  & 
  +0.107m4  & 
  -0.361m2  & 
  +0.720m2  & 
  -0.555m2  & 
  +0.142m1    
   \\[1mm]
  0.10 & 0.15 &
  -0.155m6  & 
  -0.155m6  & 
  -0.336m4  & 
  -0.234m4  & 
  +0.681m6  & 
  +0.681m6  & 
  -0.308m5  & 
  -0.216m5    
   \\[1mm]
  0.10 & 1.00 &
  -0.637m2  & 
  -0.637m2  & 
  -0.499m2  & 
  -0.497m2  & 
  +0.640m4  & 
  +0.640m4  & 
  -0.157m2  & 
  -0.156m2    
   \\[1mm]
  0.10 & 10.00 &
  -0.262p3  & 
  -0.262p3  & 
  -0.558p1  & 
  -0.558p1  & 
  -0.120p4  & 
  -0.120p4  & 
  +0.168p0  & 
  +0.168p0    
   \\[1mm]
  1.00 & 0.10 &
  +0.270m1  & 
  -0.481m1  & 
  -0.178m2  & 
  +0.172m2  & 
  -0.271m3  & 
  +0.483m3  & 
  +0.111m2  & 
  +0.278m1    
  \\[1mm]
  1.00 & 0.95 &
  +0.254m2  & 
  -0.454m2  & 
  -0.104m3  & 
  -0.258m3  & 
  -0.591m2  & 
  +0.106m1  & 
  -0.788m2  & 
  -0.242m2    
  \\[1mm]
  1.00 & 1.05 &
  -0.699m4  & 
  -0.707m4  & 
  -0.136m2  & 
  -0.311m3  & 
  +0.159m3  & 
  +0.161m3  & 
  -0.207m3  & 
  -0.464m4    
  \\[1mm]
  1.00 & 10.00 &
  -0.256p3  & 
  -0.256p3  & 
  -0.548p1  & 
  -0.548p1  & 
  -0.118p4  & 
  -0.118p4  & 
  +0.164p0  & 
  +0.164p0    
  \\[1mm]
  10.00 & 0.10 &
  +0.661m1  & 
  -0.257m1  & 
  -0.100m1  & 
  +0.375m2  & 
  +0.304p0  & 
  -0.118p0  & 
  +0.503m4  & 
  +0.187m4    
  \\[1mm]
  10.00 & 1.00 &
  +0.675p0  & 
  -0.262p0  & 
  -0.102p0  & 
  +0.378m1  & 
  +0.310p1  & 
  -0.120p1  & 
  +0.524m3  & 
  +0.194m3    
  \\[1mm]
  10.00 & 9.95 &
  +0.694m2  & 
  -0.311m2  & 
  +0.349m1  & 
  -0.321m2  & 
  -0.174m4  & 
  +0.778m5  & 
  -0.231m2  & 
  -0.330m4    
   \\[1mm]
  10.00 & 10.05 &
  -0.222m1  & 
  -0.246m1  & 
  -0.119m1  & 
  -0.267m2  & 
  -0.555m4  & 
  -0.615m4  & 
  -0.728m2  & 
  -0.265m3    
  \\[1mm]
  \hline
\end{array}
$$
}


\caption[a]{\small
 The coefficients from \eq\nr{ABCD} at a few sample points
 from the first quadrant of the $(p,p^{ }_0)$--plane
 (cf.\ \fig\ref{fig:kinematics}). 
 We have employed the notation $mX\equiv 10^{-X}_{ }$, 
 $pX \equiv 10^{+X}_{ }$. The coefficients are seen to vary by many
 orders of magnitude, implying that the interpolation described
 in \se\ref{se:tabulation} cannot reproduce all digits
 point-by-point, however it is quite accurate on the average (cf.\ the text). 
 }
\label{table:coeffs}
\end{table}
%

%
\section{Conclusions}
\la{se:conclu}

We have considered
double-differential rates for the 
production ($\Psi$, cf.\ \eq\nr{def_Psi}), 
annihilation ($\widetilde\Psi$, cf.\ \eq\nr{balance_1}), 
and scattering ($\Theta$, cf.\ \eq\nr{def_Theta})
of neutrinos and antineutrinos interacting with 
an MeV-temperature QED plasma. One application of 
such rates is that after a suitable phase-space average, 
they can be related to 
energy transfer rates between the QED and
neutrino ensembles (cf.\ \eqs\nr{edot_gain}--\nr{edot_scat}), 
which in turn parametrize averaged kinetic equations, 
describing how neutrinos decouple from the QED ensemble 
in the early universe~\cite{mea}. 
The dynamics of neutrino decoupling establishes the value
of the parameter $N_\rmi{eff}$ (cf.\ \eq\nr{def_Neff}), 
which serves as an important test of the standard cosmological model, 
as well as a strong constraint 
on its possible modifications through BSM physics. 

Specifically, we have 
evaluated $\Psi$, $\widetilde \Psi$, and $\Theta$ 
at next-to-leading order (NLO) in QED, i.e.\ including the 
full set of corrections of $\rmO(\alpha^{ }_\rmi{em})$, 
within the temperature domain $T^{ }_\gamma \ge m^{ }_e$.
This progress was possible via the quantum-mechanical
derivation of Green's functions determining
$\Psi$, $\widetilde \Psi$, and $\Theta$
(cf.\ \se\ref{se:qm}).
Once the Green's functions had been identified, they
were seen to correspond to an integrand of another 
quantity, the neutrino interaction rate, which 
had been computed up to NLO in previous work~\cite{rate}. 
Thereby, the results of ref.~\cite{rate} turned out to have
a wider range of applicability than had been originally foreseen. 

As far as the results go, 
we find NLO QED 
corrections of the expected size, i.e.\ 
on the level of $\alpha^{ }_\rmi{em} \sim 1\%$ (cf.\ \fig\ref{fig:Q}).
These corrections are in general smaller than 
those from the Maxwell-Boltzmann approximated evaluation
of leading-order energy transfer rates.
Inserting our coefficients in the code developed 
in refs.~\cite{mea2,mea}
(with a single neutrino temperature put in by hand, in order 
to reflect the effect of neutrino oscillations which were not tracked), 
and setting the electron mass to zero, we find 
$N^{ }_\rmi{eff} = 3.04858$ with the improved 
Maxwell-Boltzmann approximation, 
$N^{ }_\rmi{eff} = 3.04859$ with full leading-order rates, 
and 
$N^{ }_\rmi{eff} = 3.04867$ with NLO rates. So, NLO effects
influence the {\em fourth} decimal of $N^{ }_\rmi{eff}$.
On this point, our conclusions differ from those in ref.~\cite{cemp}, 
who had carried out a partial NLO computation within the 
Maxwell-Boltzmann approximation, but agree with ref.~\cite{new}, 
who had computed one among the many NLO diagrams. 

Another interesting small effect is the dependence of the
rates on the electron mass. Doing this at full 
{\em leading order} (cf.\ appendix~\ref{app:B}), given that NLO results
are not available with $m^{ }_e > 0$, the result 
$N^{ }_\rmi{eff} = 3.04859$ from above
gets reduced to $N^{ }_\rmi{eff} = 3.04510$,
with much of the effect originating from 
the special V$-$A contribution in \eq\nr{extra_A}. 
Therefore, the electron-mass dependence 
does influence the third decimal of $N^{ }_\rmi{eff}$.
Even though the numerical value agrees with that cited
in ref.~\cite{mea}, it appears worthwhile to explore this sensitivity
in more detail, and we plan to return to this in future work.

Beyond averaged kinetic equations, it seems 
interesting to aim at a full NLO QED estimate of $N_\rmi{eff}$.
This requires solving a set of non-averaged kinetic equations. 
Hoping that this challenge can be attacked one day, we have  
provided a tabulation and an efficient interpolation routine for 
all our double-differential rates (cf.\ \se\ref{se:tabulation}
and ref.~\cite{interpolation_code}).

%
\section*{Acknowledgements}

We thank Dietrich B\"odeker and Miguel Escudero for helpful 
discussions. 
G.J.\ was funded by the Agence Nationale de la Recherche (France), 
under grant ANR-22-CE31-0018 (AUTOTHERM), and  
benefitted also from the visitor programs 
at the University of Bern, in May 2024, 
and at CERN, in July 2024.

%
\appendix
\renewcommand{\thesection}{\Alph{section}}
\renewcommand{\thesubsection}{\Alph{section}.\arabic{subsection}}
\renewcommand{\theequation}{\Alph{section}.\arabic{equation}}

%
\section{Real and imaginary parts of the 1-loop thermal photon self-energy}
\la{app:A}

In order to make our presentation self-contained, we reproduce here 
the well-known expressions for the real and imaginary parts of 
the 1-loop photon self-energy correction, $V^{\mu\bar\mu}_{\P}$, 
in the limit of massless electrons ($m^{ }_e \ll \pi T^{ }_\gamma$). 

Because the QED plasma defines a special frame, the self-energy
is not Lorentz-covariant, and contains more structure 
than in vacuum. It can be parametrized with two separate scalar functions, 
identified with the subscripts $(...)^{ }_\rmii{T}$ and 
$(...)^{ }_\rmii{L}$ in \eq\nr{decomposition}.
However, the analytic expressions are simpler in another basis, 
identified with the subscripts 
$
 (...)^{ }_\rmii{V}
 \equiv
 -\eta^{ }_{\mu\nu}(...)^{\mu\nu}_{ }
$
and 
$
 (...)^{ }_\rmii{00}
 \equiv
 (...)^\rmii{00}_{ }
$. 
For a general transverse quantity $X$, 
the basis transformation goes as 
\be
 X^{ }_\rmii{L} \;\equiv\; - \frac{\P^2_{ }}{p^2_{ }} X^{ }_\rmii{00}
 \;, \quad
 X^{ }_\rmii{T} \;\equiv\; - \frac{X^{ }_\rmiii{V} + X^{ }_\rmiii{L}}{2} 
 \;.
 \la{basis_trafo}
\ee
This applies both to 
$ \im V^{\mu\bar\mu}_{\P} $, 
for which $X\to \rho$, and
$ \re V^{\mu\bar\mu}_{\P} $, 
for which $X\to \chi$. 

Introducing the abbreviations
\be
 \lnf(\omega) \; \equiv \; \ln \Bigl( 1 + e^{-\omega/T_\gamma} \Bigr)
 \;, \quad\;
 \lif(\omega) \;\, \equiv \; \mbox{Li}^{ }_2 \Bigl(-e^{-\omega/T_\gamma}\Bigr)
 \;, \quad\; 
 \ltf(\omega) \;\, \equiv \; \mbox{Li}^{ }_3 \Bigl(-e^{-\omega/T_\gamma}\Bigr)
 \;, \la{polylogs}
\ee
$\im V^{\mu\bar\mu}_\P$  
is parametrized via
\ba
 \rho_\rmii{V}^\rmii{LO}
  & \stackrel{p^{ }_+ > 0}{=} & 
  \frac{ \P^2_{ } }{4 \pi p }
  \Bigl\{
   p\, \theta(p^{ }_{-})  
 + 2 T^{ }_\gamma \bigl[\, \lnf(p^{ }_{+}) - \lnf(|p^{ }_{-}|) \,\bigr]
  \Bigr\}
 \;, \la{rhoV} \\[3mm] 
 \rho_\rmii{00}^\rmii{LO}
 & \stackrel{p^{ }_+ > 0}{=} & 
  \frac{ 1 }{12 \pi p }
  \Bigl\{ 
    p^3_{ }\theta(p^{ }_{-})
  + 12 p T^2_\gamma \bigl[\, \lif(p^{ }_{+})
        + \sign(p^{ }_{-})\, \lif(|p^{ }_{-}|) \,\bigr]
  \nn[3mm] & + & 
  \, 24 T^3_\gamma \bigl[\, \ltf(p^{ }_{+}) - \ltf(| p^{ }_{-} |) \,\bigr]
  \Bigr\} 
 \;. \la{rho00}
\ea
The limits for $p,|p^0_{ }| \ll \pi T^{ }_\gamma$, needed in 
\eq\nr{resum}, read
\ba
 \rho_\rmii{T}^\rmii{IR}
 & \equiv & 
 [\, \rho_\rmii{T}^\rmii{LO} \,]^{ }_{p,|p^0_{ }| \ll \pi T_\gamma}
 \; = \; 
 + \frac{\pi p^0_{ }\P^2_{ } T^2_\gamma}{12 p^3_{ }}
 \,\theta(p - |p^0_{ }|)
 \;, \la{rhoTir} \\[2mm]
 \rho_\rmii{L}^\rmii{IR}
 & \equiv & 
 [\, \rho_\rmii{L}^\rmii{LO} \,]^{ }_{p,|p^0_{ }| \ll \pi T_\gamma}
 \; = \; -
 \frac{\pi p^0_{ }\P^2_{ } T^2_\gamma}{6 p^3_{ }}
 \,\theta(p - |p^0_{ }|)
 \;. \la{rhoLir}
\ea
The values for $p^{ }_+ < 0$ can be derived 
via antisymmetry in $p^{ }_0 \to -p^{ }_0$, 
which is explicit in \eqs\nr{rhoTir} and \nr{rhoLir}.

As for $\re V^{\mu\bar\mu}_\P$, its vacuum part contains
a divergence, given in \eq\nr{chi_vac}. The thermal parts, 
separated according to \eq\nr{chi_splitup}, read
\ba
 \chi^\rmii{LO}_\rmii{V} \bigr|^{ }_{\T_{\!\gamma}}
 & = & 
 4 \int_\vec{q}
 \frac{\nF^{ }(q)}{q}
 \biggl\{\;
  -2 - \frac{\P^2_{ }}{4 p q}
 \ln \biggl|
       \frac{ 1 - [(p+2q)/p^0_{ }]^2 }
            { 1 - [(p-2q)/p^0_{ }]^2 } 
     \biggr| 
 \;\biggr\}
 \;, \la{chiV} \\[3mm] 
 \chi^\rmii{LO}_\rmii{00} \bigr|^{ }_{\T_{\!\gamma}}
 & = & 
 4 \int_\vec{q}
 \frac{\nF^{ }(q)}{q}
 \biggl\{\;
  1  + \frac{\P^2_{ } + 4 q^2}{8 p q}
 \ln \biggl|
       \frac{ 1 - [(p+2q)/p^0_{ }]^2 }
            { 1 - [(p-2q)/p^0_{ }]^2 } 
     \biggr| 
  - \frac{p^0_{ }}{2p}
 \ln \biggl|
       \frac{ 1 - [2q/(p-p^0_{ })]^2 }
            { 1 - [2q/(p+p^0_{ })]^2 } 
     \biggr| 
 \;\biggr\}
 \;. \nn \la{chi00}
\ea
The limits at $p,|p^0_{ }| \ll \pi T^{ }_\gamma$ read
\ba
 \chi_\rmii{T}^\rmii{IR}
 & \equiv & 
 [\, \chi_\rmii{T}^\rmii{LO} \,]^{ }_{p,|p^0_{ }| \ll \pi T_\gamma}
 \; = \; + 
 \frac{T^2_\gamma}{6 p^2_{ }}
 \biggl[\,  
   (p^0_{ })^2_{ }
 - \frac{p^0_{ } \P^2_{ }}{2 p}
   \ln\biggl| \frac{p+p^0_{ }}{p-p^0_{ }} \biggr|
 \,\biggr]
 \;, \la{chiTir} \\[2mm] 
 \chi_\rmii{L}^\rmii{IR}
 & \equiv & 
 [\, \chi_\rmii{L}^\rmii{LO} \,]^{ }_{p,|p^0_{ }| \ll \pi T_\gamma}
 \; = \; - 
 \frac{\P^2_{ }T^2_\gamma}{3 p^2_{ }}
 \biggl[\,  
 1 
 - \frac{p^0_{ }}{2 p}
   \ln\biggl| \frac{p+p^0_{ }}{p-p^0_{ }} \biggr|
 \,\biggr]
 \;. \la{chiLir}
\ea
The real parts are symmetric in $p^0_{ }\to -p^0_{ }$.

The limiting IR values, given in \eqs\nr{rhoTir}, \nr{rhoLir}, 
\nr{chiTir} and \nr{chiLir}, play a role in the resummation factors.
They are defined as  
\be
 \mathcal{R}^*_\rmii{T,L} 
 \; \equiv \;  
 \re \biggl\{ \frac{1}{\P^2_{ } - e^2_{ } (\chi^\rmiii{IR}_\rmiii{T,L} +
 i \rho^\rmiii{IR}_\rmiii{T,L} )}
 \biggr\}
 \; = \; 
 \frac{\P^2_{ } - e^2_{ } \chi^\rmiii{IR}_\rmiii{T,L}}
 {(\P^2_{ } - e^2_{ } \chi^\rmiii{IR}_\rmiii{T,L})^2_{ }
 + (e^2_{ } \rho^\rmiii{IR}_\rmiii{T,L})^2_{ }
 } 
 \;.
 \la{resum}
\ee

%
\section{Electron mass effects at leading order}
\la{app:B}

If the electron mass is non-zero, the vector and axial currents
that appear as parts of the operator $\mathcal{O}^\mu_{ }$, defined
in \eq\nr{H_I_full}, give independent contributions. This means 
that if we consider the corresponding Wightman function, as it appears
for the $s$-channel in \eq\nr{wight_1} and for the $t$-channel 
in \eq\nr{res_t_1}, we find
\ba
 \Pi^{\mu\bar\mu,<}_{\P}
 & \equiv & 
 \int_{\X}
 \bigl\langle\,
   \mathcal{O}^{\bar\mu}_{ }(0) \, 
   \mathcal{O}^{\mu}_{ }(\X) 
 \,\bigr\rangle^{ }_{\T_{\!\gamma}}
 \, e^{i \P\cdot\X }_{ }
 \nn[2mm]
 & = & 
 \frac{G_\rmii{F}^2 f^{ }_\rmiii{B}(p^0_{ })}{4}
 \im \biggl[\; 
  \bigl( 2\delta^{ }_{a,e} - 1 + 4 \xW \bigr)^2_{ }\; V^{\mu\bar\mu}_{\P}
  + A^{\mu\bar\mu}_{\P}
 \nn[2mm]
 & & \quad 
  \;+\,
    \bigl( 2\delta^{ }_{a,e} - 1 + 4 \xW \bigr)
    \bigl( 1 - 2\delta^{ }_{a,e} \bigr) M^{\mu\bar\mu}_{\P}
 \;\biggr]
 + \rmO(e^2_{ })
 \;. \la{wight_e}
\ea 
The vector, axial, and mixed retarded
correlators can be compactly expressed 
as the analytic continuations, 
$V^{\mu\bar\mu}_{\P}
 \equiv
 V^{\mu\bar\mu}_{P} 
 |^{ }_{p^{ }_n\to -i (p^0_{ }+ i 0^+_{ })}
$,
of the corresponding imaginary-time correlators,
\ba
 V^{\mu\bar\mu}_{P} 
 & \equiv &
 \int_X e^{i P \cdot X}_{ }
 \bigl\langle\, 
   (\bar{\ell}^{ }_e \gamma^\mu_{ } \ell^{ }_e)(X)\,
   (\bar{\ell}^{ }_e \gamma^{\bar\mu}_{ } \ell^{ }_e)(0)
 \,\bigr\rangle^{ }_{\T_{\!\gamma}}
 \;, \la{Vmunu} \\ 
 A^{\mu\bar\mu}_{P} 
 & \equiv &
 \int_X e^{i P\cdot X}_{ }
 \bigl\langle\, 
   (\bar{\ell}^{ }_e \gamma^\mu_{ } \gamma^{ }_5 \ell^{ }_e)(X)\,
   (\bar{\ell}^{ }_e \gamma^{\bar\mu}_{ } \gamma^{ }_5 \ell^{ }_e)(0)
 \,\bigr\rangle^{ }_{\T_{\!\gamma}}
 \;, \la{Amunu} \\ 
 M^{\mu\bar\mu}_{P} 
 & \equiv &
 \int_X e^{i P\cdot X}_{ }
 \bigl\langle\, 
   (\bar{\ell}^{ }_e \gamma^\mu_{ } \ell^{ }_e)(X)\,
   (\bar{\ell}^{ }_e \gamma^{\bar\mu}_{ } \gamma^{ }_5 \ell^{ }_e)(0)
 \; + \; 
   (\bar{\ell}^{ }_e \gamma^\mu_{ } \gamma^{ }_5 \ell^{ }_e)(X)\,
   (\bar{\ell}^{ }_e \gamma^{\bar\mu}_{ } \ell^{ }_e)(0)
 \,\bigr\rangle^{ }_{\T_{\!\gamma}}
 \;. \hspace*{6mm} \la{Mmunu}
\ea
Here $X \equiv (\tau,\vec{x})$, 
with $\tau \in (0,1/T^{ }_\gamma)$, 
and $P \equiv (p^{ }_n,\vec{p})$, 
with $p^{ }_n = 2\pi n T^{ }_\gamma$ and $n\in\mathbbm{Z}$. 
The overall imaginary part in \eq\nr{wight_e} refers to 
$
 \im f(p^0_{ }) \equiv 
 [f(p^0_{ } + i 0^+_{ }) - f(p^0_{ } - i 0^+_{ })]/(2 i)
$, 
i.e.\ to the discontinuity of the retarded correlator
across the real axis.   

By carrying out Wick contractions and inserting free propagators, 
it can be verified that the vector, axial, and mixed correlators
from \eqs\nr{Vmunu}--\nr{Mmunu} have the following properties: 
\bi

\item
If we contract $V^{\mu\bar\mu}_P$ with $P^{ }_\mu$, many cancellations
take place, and we find a remainder that is independent of $P$. Therefore
the imaginary part, defined as a discontinuity, vanishes. Consequently, 
$\im V^{\mu\bar\mu}_{\P}$ is transverse, and can be decomposed
like in \eq\nr{decomposition}.

\item
In contrast, the axial correlator is not transverse. However, it differs
from the vector correlator only through a single term, 
\be
 \im A^{\mu\bar\mu}_{\P}
 \; = \; 
 \im V^{\mu\bar\mu}_{\P}
 + 8 m_e^2\, \eta^{\mu\bar\mu}_{ } \, \rho^{ }_1 
 \;,  \la{extra_A}
\ee
where $\rho^{ }_1$ is defined in \eq\nr{def_rho_1}.

\item
For the mixed correlator, a Dirac trace (with a naively 
anticommuting $\gamma^{ }_5$) yields the Matsubara sum-integral
\be
 M^{\mu\bar\mu}_{P} 
 \; = \; 
 -8 i \epsilon^{\alpha\mu\beta\bar\mu}_{ }
 \int_X \Tint{\{Q,R\}}
 e^{i(P-Q+R)\cdot X}_{ }
 \frac{Q^{ }_\alpha R^{ }_\beta}{(Q^2_{ } + m_e^2 )(R^2_{ }+ m_e^2)}
 \;.
\ee
Substituting $Q\to -R$, $R\to -Q$, we see that the sum-integral
is symmetric in $\alpha\leftrightarrow\beta$. After the 
contraction with the antisymmetric Levi-Civita symbol,
$M^{\mu\bar\mu}_P$ vanishes. 

\ei

\noindent
The vanishing of $M^{\mu\bar\mu}_P$ implies that 
$\Pi^{\mu\bar\mu,<}_{\P}$ is symmetric under
$\mu\leftrightarrow\bar\mu$. Therefore, we can continue
to use the symmetrized projector 
$
 L_{\mu\bar\mu}^{ }(\K,\Q)
$ from \eq\nr{def_L} for $m^{ }_e\neq 0$.

Next, we need to carry out the Matsubara sums. Two independent 
structures are needed, 
\ba
 \rho^{ }_1
 & \equiv &
 \im  \biggl\{ 
 \int_X \Tint{\{Q,R\}}
 e^{i(P-Q+R)\cdot X}_{ }
 \frac{1}{(Q^2_{ } + m_e^2 )(R^2_{ }+ m_e^2)}
 \biggr\}^{ }_{p^{ }_n\to -i [p_0^{ }+ i 0^+_{ }]}
 \nn[2mm]
 & = & 
 \int_{\vec r} \frac{\pi}{4 \epsilon^{ }_r \epsilon^{ }_{pr}} 
 \Bigl\{\; 
  \delta(p^{ }_0 - \epsilon^{ }_r - \epsilon^{ }_{pr})
  \, \bigl[\, 1 - f^{ }_\rmii{F}(\epsilon^{ }_r)
                - f^{ }_\rmii{F}(\epsilon^{ }_{pr}) \,\bigr]
 \nn[2mm]
 & &
 \; + \, 
 2 \delta(p^{ }_0 - \epsilon^{ }_r + \epsilon^{ }_{pr})
  \, \bigl[\,     f^{ }_\rmii{F}(\epsilon^{ }_r)
                - f^{ }_\rmii{F}(\epsilon^{ }_{pr}) \,\bigr]
 \;\Bigr\}
 \;, \la{def_rho_1} \\[2mm]
 \rho^{ }_2
 & \equiv &
 \im  \biggl\{ 
 \int_X \Tint{\{Q,R\}}
 e^{i(P-Q+R)\cdot X}_{ }
 \frac{q^{ }_n r^{ }_n}{(Q^2_{ } + m_e^2 )(R^2_{ }+ m_e^2)}
 \biggr\}^{ }_{p^{ }_n\to -i [p_0^{ }+ i 0^+_{ }]}
 \nn[2mm]
 & = & 
 \int_{\vec r} \frac{\pi}{4} 
 \Bigl\{\; 
  \delta(p^{ }_0 - \epsilon^{ }_r - \epsilon^{ }_{pr})
  \, \bigl[\, 1 - f^{ }_\rmii{F}(\epsilon^{ }_r)
                - f^{ }_\rmii{F}(\epsilon^{ }_{pr}) \,\bigr]
 \nn[2mm]
 & &
 \; - \, 
 2 \delta(p^{ }_0 - \epsilon^{ }_r + \epsilon^{ }_{pr})
  \, \bigl[\,     f^{ }_\rmii{F}(\epsilon^{ }_r)
                - f^{ }_\rmii{F}(\epsilon^{ }_{pr}) \,\bigr]
 \;\Bigr\}
 \;, \la{def_rho_2}
\ea
where $\epsilon^{ }_r \equiv \sqrt{r^2_{ }+ m_e^2}$ 
and $\epsilon^{ }_{pr} \equiv \sqrt{(\vec{p+r})^2_{ }+ m_e^2}$.
Subsequently, angular integrations can be performed. Denoting
\be
 \epsilon^{ }_{\pm} 
 \; \equiv \;
 \frac{p^{ }_0 \pm p \sqrt{\Delta}}{2}
 \;, \quad
 \Delta
 \; \equiv \; 
 \biggl( 1 - \frac{4 m_e^2}{\P^2_{ }} \biggr)
 \theta
 \biggl( 1 - \frac{4 m_e^2}{\P^2_{ }} \biggr)
 \;, \la{e_pm}
\ee
and generalizing \eq\nr{spectral_int} into
\ba
\langle ... \rangle
& \equiv &
\frac{1}{16 \pi p}
 \biggl\{ 
   \theta(\P^2_{ } - 4 m_e^2)
   \int_{\epsilon^{ }_-}^{ \epsilon^{ }_+ } \! {\rm d}r^{ }_0  
  - \theta(-\P^2_{ })
 \biggl[  \int_{-\infty}^{ \epsilon^{ }_-}
        + \int_{\epsilon^{ }_+}^{\infty} \biggr] 
 \, {\rm d}r^{ }_0
 \biggr\} 
 \nn[2mm] 
 & & \; \times \, 
 \bigl[ 1 - \nF^{ }(p^{ }_0 - r^{ }_0) - \nF^{ }(r^{ }_0) \bigr]
 (...)
 \;,
 \la{spectral_int_e}
\ea
we find 
\be
 \rho^{ }_1 \; = \; \langle 1 \rangle
 \;, \quad
 \rho^{ }_2 \; = \; \langle r^{ }_0 (p^{ }_0 - r^{ }_0) \rangle
 \;. 
\ee
Finally, we can integrate over $r^{ }_0$. With 
the notation from \eqs\nr{polylogs} and \nr{e_pm}, this yields
\ba
 \rho^{ }_1
  & = & 
  \frac{ 1 }{16 \pi p }
  \Bigl\{\, 
   \theta(\P^2_{ } - 4m_e^2)\,
   p\, \sqrt{\Delta}  
 + 2 T^{ }_\gamma
 \bigl[\, \lnf(\epsilon^{ }_{+}) - \lnf(|\epsilon^{ }_{-}|) \,\bigr]
  \,\Bigr\}
 \;, \la{rho_1} \\[3mm] 
 \rho^{ }_2
 & = & 
  \frac{ 1 }{16 \pi p }
  \Bigl\{\, 
   \theta(\P^2_{ } - 4m_e^2)\,
   \frac{p\sqrt{\Delta}}{4}
   \biggl( p_0^2 - \frac{p^2_{ }\Delta }{3} \biggr)
  + \frac{(p_0^2 - p^2_{ }\Delta)T^{ }_\gamma}{2}
  \bigl[\, \lnf(\epsilon^{ }_{+}) - \lnf(|\epsilon^{ }_{-}|) \,\bigr]
 \nn[2mm]
 & & \; + \,
   2 p \sqrt{\Delta}\, T^2_\gamma
           \bigl[\, \lif(\epsilon^{ }_{+})
        + \sign(\epsilon^{ }_{-})\, \lif(|\epsilon^{ }_{-}|) \,\bigr]
   +  
  \, 4 T^3_\gamma \bigl[\, \ltf(\epsilon^{ }_{+})
   - \ltf(| \epsilon^{ }_{-} |) \,\bigr]
  \,\Bigr\} 
 \;. \la{rho_2}
\ea

Now, we can put the results together. Contracting the vector structure 
from \eq\nr{decomposition}, or the additional axial contribution from 
\eq\nr{extra_A}, with $L^{ }_{\mu\bar\mu}$ from \eq\nr{def_L}, we can 
represent the double-differential rates $\Psi$ and $\Theta$ in terms
a common function $\F$, like in \eqs\nr{Psi_NLO} and \nr{Theta_NLO}. 
This yields a generalization of \eqs\nr{Psi_LO} and \nr{Theta_LO}
to the massive situation,  
\ba
 \F 
 & = & 
 \P^2_{ } \, \biggl\{\, 
  \Bigl[
    \bigl( 
      2\delta^{ }_{a,e} - 1 + 4 \xW
    \bigr)^2_{ } + 1 
  \Bigr]
  \biggl[\, 
    \rho^\rmii{LO}_\rmii{T} + \rho^\rmii{LO}_\rmii{L} 
  + 
   \biggl( \frac{2 k - p^{ }_0}{p} \biggr)^2_{ }
   \bigl(\, \rho^\rmii{LO}_\rmii{T} - \rho^\rmii{LO}_\rmii{L} \,\bigr)
  \,\biggr]
 \nn[2mm]
 &  & \;+\, 
 16\; m_e^2 \; \rho^{ }_1 
 \,\biggr\}
 \; + \; \rmO(e^2_{ })
 \;.  \la{calF_e}
\ea
Here $\rho^\rmii{LO}_\rmii{T}$ and $\rho^\rmii{LO}_\rmii{L}$ are 
represented like in \eq\nr{basis_trafo}, 
\be
 \rho^\rmii{LO}_\rmii{L}
  \;\equiv\;
  - \frac{\P^2_{ }}{p^2_{ }} \rho^\rmii{LO}_\rmii{00}
 \;, \quad
 \rho^\rmii{LO}_\rmii{T}
  \;\equiv\;
  - \frac{\rho^\rmii{LO}_\rmiii{V} + \rho^\rmii{LO}_\rmiii{L}}{2}  
 \;, 
\ee
where in turn 
\ba 
 \rho_\rmii{V}^\rmii{LO}
 & \equiv & 
 -\eta^{ }_{\mu\bar\mu} \im V^{\mu\bar\mu}_{\P}
 \; = \;
 4 \,\bigl(\, \P^2_{ } + 2 m_e^2 \bigr) \, \rho^{ }_1 
 \;, \\[2mm]
 \rho_\rmii{00}^\rmii{LO} 
 & = & 
 \im V^{00}_{\P}
 \; = \; 
 2 \,\bigl(\,
 - \P^2_{ } \, \rho^{ }_1 
 + 4 \, \rho^{ }_2 
 \,\bigr)
 \;.
\ea
The spectral functions $\rho^{ }_1$ and $\rho^{ }_2$ are given 
in \eqs\nr{rho_1} and \nr{rho_2}, respectively. 

Given the structure in \eq\nr{calF_e}, 
we generalize \eq\nr{calF_rep} to 
\be
 \mathcal{F} 
 \; \equiv \; 
 T^4_{\gamma} \, 
 \biggl\{\, 
   \widehat{\mathcal{F}}^{ }_\rmii{T} 
  + 
   \widehat{\mathcal{F}}^{ }_\rmii{L} 
  + 
   \biggl( \frac{2k - p^{ }_0}{p} \biggr)^2_{ } 
   \bigl(
   \widehat{\mathcal{F}}^{ }_\rmii{T} 
  -
   \widehat{\mathcal{F}}^{ }_\rmii{L} 
   \bigr)
 + \widehat{\mathcal{F}}^{ }_\rmii{1} 
 \,\biggr\}
 \;, \la{calF_rep_e}
\ee
and \eq\nr{ABCD} to 
\be
 \widehat{\mathcal{F}}^{ }_\rmii{T,L}
 \; 
 \equiv 
 \;
 \bigl[ \,
   \bigl( 2\delta^{ }_{a,e} - 1 + 4 \xW \bigr)^2_{ }
 + 1 
 \, \bigr] \,
   \underbrace{ 
   \biggl( \frac{\P^2_{ } \rho^\rmii{LO}_\rmiii{T,L}}{T^4_{\gamma}} \biggr)
   }_{ 
   \,\equiv\, \A^{ }_\rmiii{T,L} }
   \; + \; \rmO(e^2_{ })
 \;, 
 \quad
 \widehat{\mathcal{F}}^{ }_\rmii{1}
 \;
 \equiv 
 \; 
   \underbrace{ 
  \frac{16 m_e^2 \P^2_{ } \rho^{ }_1}{T_\gamma^4}
   }_{ 
   \,\equiv\, \A^{ }_\rmiii{1} }
   \; + \; \rmO(e^2_{ }) 
 \;. \la{ABCD_e_2}
\ee
The results for the coefficients $\A^{ }_\rmii{T}$, 
$\A^{ }_\rmii{L}$, $\A^{ }_\rmii{1}$
are illustrated in table~\ref{table:coeffs_e}.

%
\begin{table}[t]

{\fontsize{8pt}{10pt}\selectfont
$$
\begin{array}{rrrrrrrr} 
  & &
  \multicolumn{3}{|c|}{ m^{ }_e/T^{ }_\gamma = 1.0 } 
  & 
  \multicolumn{3}{|c|}{ m^{ }_e/T^{ }_\gamma = 5.0 } 
 \\[1mm]
  p / T_\gamma^{ } & p_0^{ } / T_\gamma^{ }
  & \mathcal{A}^{ }_\rmii{T} 
  & \mathcal{A}^{ }_\rmii{L} 
  & \mathcal{A}^{ }_\rmii{1} 
  & \mathcal{A}^{ }_\rmii{T} 
  & \mathcal{A}^{ }_\rmii{L} 
  & \mathcal{A}^{ }_\rmii{1} 
 \\[1mm]
  \hline \\[-3mm]
  0.10 & 0.05 &
  +0.546m3  & 
  -0.138m2  & 
  +0.572m3  & 
  +0.188m4  & 
  -0.130m3  & 
  +0.185m3    
   \\[1mm]
  0.10 & 0.15 &
  +0.0p0 & 
  +0.0p0  & 
  +0.0p0  & 
  +0.0p0  & 
  +0.0p0  & 
  +0.0p0    
   \\[1mm]
  0.10 & 1.00 &
  +0.0p0  & 
  +0.0p0  & 
  +0.0p0  & 
  +0.0p0  & 
  +0.0p0  & 
  +0.0p0    
   \\[1mm]
  0.10 & 10.00 &
  -0.261p3  & 
  -0.261p3  & 
  +0.308p2  & 
  +0.0p0  & 
  +0.0p0  & 
  +0.0p0    
   \\[1mm]
  1.00 & 0.10 &
  +0.212m1  & 
  -0.463m1  & 
  +0.155m1  & 
  +0.126m2  & 
  -0.743m2  & 
  +0.100m1    
  \\[1mm]
  1.00 & 0.95 &
  +0.514m3  & 
  -0.212m2  & 
  +0.229m2  & 
  +0.568m8  & 
  -0.955m7  & 
  +0.169m6    
  \\[1mm]
  1.00 & 1.05 &
  +0.0p0  & 
  +0.0p0  & 
  +0.0p0  & 
  +0.0p0  & 
  +0.0p0  & 
  +0.0p0    
  \\[1mm]
  1.00 & 10.00 &
  -0.256p3  & 
  -0.256p3  & 
  +0.304p2  & 
  +0.0p0  & 
  +0.0p0  & 
  +0.0p0    
  \\[1mm]
  10.00 & 0.10 &
  +0.601m1  & 
  -0.256m1  & 
  +0.386m2  & 
  +0.893m2  & 
  -0.111m1  & 
  +0.135m1    
  \\[1mm]
  10.00 & 1.00 &
  +0.613p0  & 
  -0.261p0  & 
  +0.398m1  & 
  +0.897m1  & 
  -0.112p0  & 
  +0.137p0    
  \\[1mm]
  10.00 & 9.95 &
  +0.235m4  & 
  -0.787m4  & 
  +0.127m3  & 
  +0.332m21  & 
  -0.664m21  & 
  -0.0m0    
   \\[1mm]
  10.00 & 10.05 &
  +0.0p0  & 
  +0.0p0  & 
  +0.0p0  & 
  +0.0p0  & 
  +0.0p0  & 
  +0.0p0    
  \\[1mm]
  \hline
\end{array}
$$
}


\caption[a]{\small
 The coefficients from \eq\nr{ABCD_e_2} 
 at a few sample points
 from the first quadrant of the $(p,p^{ }_0)$--plane
 (cf.\ \fig\ref{fig:kinematics}). 
 We have employed the notation $mX\equiv 10^{-X}_{ }$, 
 $pX \equiv 10^{+X}_{ }$. 
 }
\label{table:coeffs_e}
\end{table}
%

\small{
%

}



\begin{thebibliography}{99}

\bibitem{Neffm2}
  K.~Akita and M.~Yamaguchi,
  {\it A precision calculation of relic neutrino decoupling,}
  JCAP {08} (2020) 012
  [\href{https://arxiv.org/abs/2005.07047}{2005.07047}].

\bibitem{Neffm1}
  J.~Froustey, C.~Pitrou and M.C.~Volpe,
  {\it Neutrino decoupling including flavour oscillations
  and primordial nucleosynthesis,}
  JCAP {12} (2020) 015
  [\href{https://arxiv.org/abs/2008.01074}{2008.01074}].

\bibitem{Neff0}
  J.J.~Bennett, G.~Buldgen, P.F.~De Salas, M.~Drewes, S.~Gariazzo, 
  S.~Pastor and Y.Y.Y.~Wong,
  {\it Towards a precision calculation of $N_{\rm eff}$ in the Standard Model.
  Part II. Neutrino decoupling in the presence of flavour oscillations
  and finite-temperature QED,}
  JCAP {04} (2021) 073
  [\href{https://arxiv.org/abs/2012.02726}{2012.02726}].

\bibitem{Neff4}
  M.~Drewes, Y.~Georis, M.~Klasen, G.~Pierobon and Y.Y.Y.~Wong,
  {\it Towards a precision calculation of $N_{\rm eff}$ 
  in the Standard Model IV: Impact of positronium formation,}
  [\href{https://arxiv.org/abs/2411.14091}{2411.14091}].
  %

\bibitem{cemp}
  M.~Cielo, M.~Escudero, G.~Mangano and O.~Pisanti,
  {\it $N^{ }_{\rm eff}$ in the Standard Model at NLO is 3.043,}
  Phys.\ Rev.\ D {108} (2023) L121301
  [\href{https://arxiv.org/abs/2306.05460}{2306.05460}].

\bibitem{rate}
  G.~Jackson and M.~Laine,
  {\it QED corrections to the thermal neutrino interaction rate,}
  JHEP {05} (2024) 089
  [\href{https://arxiv.org/abs/2312.07015}{2312.07015}].

\bibitem{new}
  M.~Drewes, Y.~Georis, M.~Klasen, L.P.~Wiggering and Y.Y.Y.~Wong,
  {\it Towards a precision calculation of $N_{\rm eff}^{ }$ in the 
  Standard Model. Part III. Improved estimate of NLO corrections
  to the collision integral,}
  JCAP {06} (2024) 032
  [\href{https://arxiv.org/abs/2402.18481}{2402.18481}].

\bibitem{mea2}
  M.~Escudero,
  {\it Neutrino decoupling beyond the Standard Model: 
  CMB constraints on the Dark Matter mass with a fast 
  and precise $N_{\rm eff}$ evaluation,}
  JCAP {02} (2019) 007
  [\href{https://arxiv.org/abs/1812.05605}{1812.05605}].

\bibitem{mea}
  M.~Escudero Abenza,
  {\it Precision early universe thermodynamics made simple: 
  $N^{ }_{\rm eff}$ and neutrino decoupling in the Standard Model
  and beyond,}
  JCAP {05} (2020) 048
  [\href{https://arxiv.org/abs/2001.04466}{2001.04466}].

\bibitem{htl6}
  E.~Braaten and R.D.~Pisarski,
  {\it Simple effective Lagrangian for hard thermal loops,}
  Phys.\ Rev.\ D {45} (1992) 1827. 

\bibitem{eks}
  K.~Enqvist, K.~Kainulainen and V.~Semikoz,
  {\it Neutrino annihilation in hot plasma,}
  Nucl.\ Phys.\ B {374} (1992) 392.

\bibitem{twoloop}
  G.~Jackson,
  {\it Two-loop thermal spectral functions with general kinematics,}
  Phys.\ Rev.\ D {100} (2019) 116019
  [\href{https://arxiv.org/abs/1910.07552}{1910.07552}].

\bibitem{twoloop_code}
  G.~Jackson,
  {\it Numerical code for master integrals for thermal spectral functions},
  \href{https://doi.org/10.5281/zenodo.3478143}{\tt https://doi.org/10.5281/zenodo.3478143}
  \ (2019). 

\bibitem{eft2}
  R.J.~Hill and O.~Tomalak,
  {\it On the effective theory of neutrino-electron
  and neutrino-quark interactions,}
  Phys.\ Lett.\ B {805} (2020) 135466
  [\href{https://arxiv.org/abs/1911.01493}{1911.01493}].

\bibitem{interpolation_code}
  G.~Jackson,
  {\it Tabulation and interpolation of NLO neutrino-antineutrino
  production and scattering rates at MeV temperatures},
  \href{https://doi.org/10.5281/zenodo.14217713}{\tt https://doi.org/10.5281/zenodo.14217713}
  \ (2024). 

\end{thebibliography}
\end{document}